\documentclass[12pt]{article}
\usepackage{amsmath,amssymb,exscale,cancel}
\usepackage{slashed}
\allowdisplaybreaks
\usepackage{graphicx}
\usepackage{graphicx}
\usepackage{epsfig}
\usepackage{multicol}
\usepackage{color}
\usepackage{mathrsfs}
\usepackage{blindtext}
 \usepackage{fancyhdr}
\usepackage{hyperref}
\usepackage{cite}
\usepackage{mathtools}

\usepackage{floatrow}

\usepackage{graphicx}
\usepackage{sidecap}

\DeclareMathOperator*{\SumInt}{%
\mathchoice%
  {\ooalign{$\displaystyle\sum$\cr\hidewidth$\displaystyle\int$\hidewidth\cr}}
  {\ooalign{\raisebox{.14\height}{\scalebox{.7}{$\textstyle\sum$}}\cr\hidewidth$\textstyle\int$\hidewidth\cr}}
  {\ooalign{\raisebox{.2\height}{\scalebox{.6}{$\scriptstyle\sum$}}\cr$\scriptstyle\int$\cr}}
  {\ooalign{\raisebox{.2\height}{\scalebox{.6}{$\scriptstyle\sum$}}\cr$\scriptstyle\int$\cr}}
}

\usepackage[latin1]{inputenc} 

\usepackage{multicol}  
 
 \usepackage{titlesec}
 
 \usepackage{rotating,slashed,xcolor,amsfonts,expdlist,charter}

\numberwithin{equation}{section}

\usepackage{xcolor}
\usepackage{sectsty}

\usepackage{mdframed}
\usepackage{titletoc}

\numberwithin{equation}{section}

\usepackage{xcolor}
\usepackage{sectsty}

\usepackage{mdframed}
\usepackage{titletoc}

\definecolor{secnum}{RGB}{13,151,225}
\definecolor{ptcbackground}{RGB}{212,237,252}
\definecolor{ptctitle}{RGB}{0,177,235}

\titlecontents{lsection}
  [5.8em]{\sffamily}
  {\color{secnum}\contentslabel{2.3em}\normalcolor}{}
  {\titlerule*[1000pc]{.}\contentspage\\\hspace*{-5.8em}\vspace*{5pt}%
    \color{white}\rule{\dimexpr\textwidth-15.5pt\relax}{1pt}}

\usepackage{hyperref}
\hypersetup{colorlinks,bookmarksopen,bookmarksnumbered,citecolor=rossos,
linkcolor=redy,pdfstartview=FitH,urlcolor=green-go}
\usepackage{slashed}

\definecolor{blus}{cmyk}{1,0.9,0,0.1}
\definecolor{verdes}{cmyk}{0.99,0,0.59,0.65}
\definecolor{rossos}{cmyk}{0,1,1,0.55}
\definecolor{redy}{cmyk}{0,1,1,0.7}
\definecolor{greeny}{cmyk}{0.99,0,0.59,0.98}
\definecolor{green-go}{cmyk}{0.79,0,0.59,0.5}

\usepackage{titlesec}

\newcommand{\beq}{\begin{equation}}
\newcommand{\eeq}{\end{equation}}

\def\hhref#1{\href{http://arxiv.org/abs/#1}{arXiv:#1}} 

\newcommand{\tmtextbf}[1]{{\bfseries{#1}}}
\newcommand{\tmtextrm}[1]{{\rmfamily{#1}}}

\def\be{\begin{equation}}
\def\ee{\end{equation}}
\def\ba{\begin{array} }
\newcommand{\Tr}{\,{\rm Tr}}
\def\bac{\begin{array} {c}}
\def\bacc{\begin{array} {cc}}
\def\baccc{\begin{array} {ccc}}
\def\bacccc{\begin{array} {cccc}}
\def\ea{\end{array}}
\def\bea{\begin{eqnarray}}
\def\eea{\end{eqnarray}}

\definecolor{red}{rgb}{1,0,0}

\def\psl{\hbox{\hbox{${p}$}}\kern-1.9mm{\hbox{${/}$}}}
\def\dsl{\hbox{\hbox{${\partial}$}}\kern-2.2mm{\hbox{${/}$}}}
\def\Dsl{\hbox{\hbox{${D}$}}\kern-2.6mm{\hbox{${/}$}}}

\newcommand{\gappeq}{{\rlap{{\raise}.5ex\text{\ensuremath{>}}}{{\lower}.5ex\text{\ensuremath{\sim}}}}}
\newcommand{\lappeq}{{\rlap{{\raise}.5ex\text{\ensuremath{<}}}{{\lower}.5ex\text{\ensuremath{\sim}}}}}
\newcommand{\I}{\tmtextrm{1{\kern}-.24em l}}

\begin{document}
\topmargin -1.0cm
\oddsidemargin 0.9cm
\evensidemargin -0.5cm

{\vspace{-1cm}}
\begin{center}

\vspace{-1cm}

 {\LARGE \tmtextbf{ 
\color{rossos} \hspace{-1.9cm}   
Scalar Thermal Field Theory for a Rotating Plasma 
 \hspace{-1.6cm}}} {\vspace{.5cm}}\\


\vspace{1.3cm}

{\large{\bf  Alberto Salvio }}

{\em  
\vspace{.4cm}
 Physics Department, University of Rome Tor Vergata, \\ 
via della Ricerca Scientifica, I-00133 Rome, Italy\\

\vspace{0.6cm}

I. N. F. N. -  Rome Tor Vergata,\\
via della Ricerca Scientifica, I-00133 Rome, Italy\\

  \vspace{0.5cm}

}
\vspace{1.5cm}
\end{center}

\noindent ---------------------------------------------------------------------------------------------------------------------------------
\begin{center}
{\bf \large Abstract}
\end{center}
\noindent   This paper initiates the systematic study of thermal field theory for generic equilibrium density matrices, which feature arbitrary values not only of temperature and chemical potentials, but also of average angular momentum. The focus here is on scalar fields, although some results also apply to fields with arbitrary spins. A general technique to compute ensemble averages is provided.  Moreover, path-integral methods are developed to study thermal Green's functions (with an arbitrary number of points) in generic theories, which cover both the real-time and imaginary-time formalism. It is shown that, while the average angular momentum, like the chemical potentials, does not contribute positively to the Euclidean action, its negative contributions can be compensated by some other contributions that are instead positive, at least in some cases, e.g.~when the chemical potentials vanish. 
As an application of the developed general formalism, it is shown that the production of particles weakly coupled to a rotating plasma can be significantly enhanced compared to the non-rotating case. 
The Higgs boson production through a portal coupling to a dark sector in the early universe is studied in some detail. The findings of this paper can also be useful, for example, to investigate the physics of rotating stars, ordinary and primordial black holes and more exotic compact objects.

\vspace{0.7cm}

\noindent---------------------------------------------------------------------------------------------------------------------------------

\newpage

\tableofcontents

\noindent --------------------------------------------------------------------------------------------------------------------------------

\vspace{0.5cm}

\section{Introduction}\label{intro}

When applying the laws of physics to cases of interest we often face difficulties due to a large number of particles. Of course, this can happen even when particles move at relativistic speeds and/or quantum effects must be taken into account, like in the early universe. In these situations one can use thermal field theory (TFT),  namely the combination of statistical mechanics and relativistic quantum field theory. Today TFT has become the standard theoretical tool to study particle physics processes (decays, scattering processes, particle production, phase transitions, etc.) in a thermal medium (see~\cite{Bellac:2011kqa,Nair:2005iw} for textbooks,~\cite{Landsman:1986uw,Quiros:1994dr,Laine:2016hma} for monographs and~\cite{Salvio:2024upo} for a recent introduction from first principles).

At (thermodynamic) equilibrium the density matrix, the key input in TFT, can be expressed in terms of all conserved quantities, which, in a relativistic setting, include the Hamiltonian, the linear and angular momentum and all conserved charges~\cite{LandauLifshitz}. Such a density matrix has been studied in a number of works (see e.g.~\cite{Zubarev:1979afm,Weert,Becattini:2012tc}). However, TFT has been systematically developed only in the absence of the angular momentum in the density matrix, i.e.~for vanishing average angular momentum (see, however,~\cite{Vilenkin:1980zv,Buzzegoli:2017cqy,Becattini:2020qol,Buzzegoli:2020ycf,Palermo:2021hlf,Palermo:2023cup} for some analysis of free scalar and Dirac fields in a rotating plasma).

The purpose of this paper is to initiate a systematic study of (generically interacting) TFT for the most general equilibrium density matrix, including not only temperature and chemical potentials associated with the conserved charges, but also a non-vanishing value of the average angular momentum. In this work we focus on scalar fields to avoid further technical complications due to higher spins\footnote{See Refs.~\cite{Salvio:2025ggj} and~\cite{Salvio:2026ewl} for a recent extension of the present work to general gauge theories coupled to an arbitrary matter sector.}.

For obvious reasons, the ensemble averages of observables are among the most important quantities that one can compute in TFT. For less obvious reasons, very important are also the thermal Green's functions, which are defined as the statistical average of the expectation values of the time-ordered product of a generic number of fields taken on a complete set of states. Indeed, the applications of thermal Green's functions include, among other things, the determination of the effective action, which allows us, for example, to study possible phase transitions, and the computation of rates of particle processes (decays, scattering processes and particle production). Therefore, the main goal of this paper is to provide systematic techniques to determine the ensemble averages of observables and the thermal Green's function for the most general equilibrium density matrix.

In the generically interacting case an approach that can give us both these quantities is the path-integral one. So  in this work the path integral representation of the partition function, which gives us the ensemble averages of observables, and of the Green's functions is investigated.   The path integral, when formulated at imaginary time, i.e.~on the Euclidean spacetime,  also offers a way to study the theory non perturbatively. As well-known, the real chemical potentials cannot be included non perturbatively in the path integral because of the sign problem. Thus, an interesting question is whether or not the average angular momentum leads to a similar issue. Another purpose of this paper is to shed light on this aspect.  

Several motivations for these studies come to mind. Their applications can include phase transitions, decays, scattering processes and particle production around rotating stars, ordinary and primordial black holes and exotic compact objects. For example, the accretion disks and coronas around black holes can be often considered rotating plasmas in approximate thermodynamic equilibrium.  Moreover, one can conceive investigating the same phenomena (phase transitions, decays, scattering processes and particle production) in a lab, engineering a rotating plasma. 

The paper is organized as follows. 
\begin{itemize}
\item In the next section the general equilibrium density matrix is expressed in a convenient reference frame where the plasma is at rest, but conversion formul\ae~are provided to recover the expressions in general inertial frames. Moreover, the ensemble averages of observables are given with no assumption on the underlying theory. 
\item As a first step towards the goals of this paper, in Sec.~\ref{Free fields} the free field case is studied, keeping, however, a general number of fields and general values of particle masses, temperature, chemical potentials and average angular momentum.  
\item Sec.~\ref{Interacting theories and general path-integral formula} is devoted to the derivation of the general path-integral formula for the partition function and the Green's functions, without committing ourselves to any specific underlying theory. However, the more explicit expressions for scalar theories commonly encountered in particle physics are worked out in Sec.~\ref{MomInt}.
\item Finally, Sec.~\ref{Applications} illustrates some applications of the general results previously obtained. In particular, that section shows how to compute the production of a spin-0 particle that is weakly coupled to the plasma for generic values of temperature, chemical potentials and average angular momentum. The main goal is to investigate how the average angular momentum  influences the particle production rate.  
\item Sec.~\ref{Conclusions} provides a detailed summary of the main original results of this paper and the final conclusions.
\end{itemize}

\section{General equilibrium density matrix }\label{density matrix}

The most general equilibrium density matrix $\rho_{\rm gen}$ is an exponential function of the Hamiltonian $H$ and the other conserved quantities, which commute with $H$. So  in the relativistic case
  \be \rho_{\rm gen}= \frac1{Z}\exp\left(-\beta_\mu P^\mu-\frac12\beta_{ij}J_{ij} - \beta_a Q^a\right), \label{EqDen}\ee 
  where $P^0=H$, the $P^i$ are the components of the (linear) momentum, $J_k \equiv  \frac12 \epsilon_{ijk}J_{ij}$ (with $J_{ij}=-J_{ji}$ and $\epsilon_{ijk}$ being the totally antisymmetric Levi-Civita symbol such that $\epsilon_{123}=1$) are the components of  the angular momentum, the $Q^a$ are the full set of charges, which generate a possible continuous internal (not necessarily Abelian) symmetry group $\mathcal{G}$, if any, and the $\beta_\mu$, $\beta_{ij}$ and $\beta_a$ are real constants. Since $J_{ij}=-J_{ji}$ we require $\beta_{ij}=-\beta_{ji}$ without loss of generality.  Also, in this work repeated indices understand a summation (unless otherwise stated) and the convention for the spacetime metric is $\eta=\,$diag$(1,-1,-1,-1)$ (its components $\eta_{\mu\nu}$ are used as usual to lower and raise the spacetime indices). The partition function $Z$ is
  \be Z = \Tr\exp\left(-\beta_\mu P^\mu-\frac12\beta_{ij}J_{ij} - \beta_a Q^a\right). \label{ZTr}\ee 
  
  \subsection{Rest frame and conversion formul\ae~}\label{Rest frame}
   We can equivalently write $\beta_{ij} J_{ij} = 2\vec\tau \cdot \vec J$, where $\tau_k\equiv \frac12 \epsilon_{ijk}\beta_{ij}$  and $\vec\tau \cdot \vec J\equiv \tau_k J_k$, so that
  \be \rho_{\rm gen}= \frac1{Z}\exp\left(-\beta_0 H-\vec\beta\cdot \vec P-\vec\tau \cdot \vec J - \beta_a Q^a\right), \label{EqDen2}\ee
  with $\vec\beta\cdot \vec P \equiv \beta_i P^i$.
  Furthermore, we can decompose $\vec\beta$ as $\vec\beta = \vec\beta^{\parallel}+\vec\beta^{\bot}$, where $\vec\beta^{\parallel}$ and $\vec\beta^{\bot}$ are, respectively, parallel and orthogonal to $\vec\tau$. The operator $\vec\beta^{\parallel}\cdot \vec P$ commutes with $\beta_0 H, \vec\beta^{\bot}\cdot \vec P, \vec\tau \cdot \vec J$ and $\beta_a Q^a$, but $\vec\beta^{\bot}\cdot \vec P$ only commutes with $\beta_0 H, \vec\beta^{\parallel}\cdot \vec P$ and $\beta_a Q^a$. 
  
  The technical complications due to the fact that $\vec\beta^{\bot}\cdot \vec P$ and $\vec\tau \cdot \vec J$ do not commute with each other can be avoided by considering a space translation of the reference frame:
  \be U(\vec a) \rho_{\rm gen} U(\vec a)^\dagger = \frac1{Z}\exp\left(-\beta_0 H-\vec\beta\cdot \vec P-\frac12\beta_{ij}(J_{ij}+a_iP_j-a_jP_i) - \beta_a Q^a\right),\ee
  where $U(\vec a) = \exp(-i a_jP^j)$ is the unitary operator representing the space translation associated with $\vec a$ on the Hilbert space and the well-known commutation rules between the linear and angular momentum and the Hamiltonian of a relativistic theory have been used. Using $\beta_{ij}=\epsilon_{ijk}\tau_k$ one obtains
   \be  U(\vec a) \rho_{\rm gen} U(\vec a)^\dagger= \frac1{Z}\exp\left(-\beta_0 H-(\vec\beta-\vec\tau\times\vec a)\cdot \vec P-\vec\tau \cdot \vec J - \beta_a Q^a\right)\ee
   and we can find $\vec a$ such that $\vec\tau\times\vec a$ is opposite to $\vec\beta^\bot$, so
   \be  U(\vec a) \rho_{\rm gen} U(\vec a)^\dagger= \frac1{Z}\exp\left(-\beta_0 H-\vec\beta^\parallel\cdot \vec P-\vec\tau \cdot \vec J - \beta_a Q^a\right). \ee
   In this new reference frame the density matrix is a function of only commuting operators. 
   
   Now, the convergence of the trace in~(\ref{ZTr})
   requires\footnote{In Ref.~\cite{Salvio:2024upo} $\vec\tau=0$ was assumed, but one can consider the terms in the trace in~(\ref{ZTr}) with vanishing eigenvalue of $\vec\tau\cdot \vec J$ and so $\beta_0>0$ and $|\beta^\parallel|<\beta_0$ remain necessary for that convergence.} $\beta_0>0$ and $|\vec\beta^\parallel|<\beta_0$~\cite{Salvio:2024upo}, meaning that $\{\beta_0, \vec\beta^\parallel\}$ form a time-like vector. So  one can now consider a Lorentz transformation $\Lambda$, implemented by a unitary operator $U(\Lambda)$ on the Hilbert space, such that $U(\Lambda)\, \vec\tau \cdot \vec J \, U(\Lambda)^\dagger=\vec\tau \cdot \vec J$ and 
    \be U(\Lambda)U(\vec a) \rho_{\rm gen} U(\vec a)^\dagger U(\Lambda)^\dagger = \frac{e^{-\beta H-\vec\tau  \cdot \vec J - \beta_a Q^a}}{Z} \equiv \rho, \label{rhoRest0}
  \ee
  where $\beta \equiv \sqrt{\beta_0^2-|\vec\beta^\parallel|^2}>0$ is the inverse temperature, $\beta\equiv1/T$. To obtain~(\ref{rhoRest0}) one can use the fact that  a Lorentz transformation that brings the time-like vector $\{\beta_0, \vec\beta^\parallel\}$ into $\{\beta, 0\}$ can involve only the component of the boost operator along $\vec\tau$, which commutes with $\vec\tau\cdot\vec J$. Sometimes $\vec\tau$  is named thermal vorticity.
  
  It is also convenient to express $\rho$ in the following form
   \be \rho = \frac{e^{-\beta (H-\vec\Omega  \cdot \vec J - \mu_a Q^a)}}{Z}, \label{rhoRest}
  \ee
  where $\mu_a\equiv - \beta_a/\beta$ is the chemical potential associated with $Q^a$ and $\vec\Omega\equiv -\vec \tau/\beta$.   The operator
$\rho$ represents here the density matrix in the frame  where\footnote{For a generic operator $\mathcal{F}$ the ensemble average is 
$\langle \mathcal{F}\rangle =  \Tr(\rho \mathcal{F})$.}
 $\langle \vec P\rangle = \Tr(\rho\vec P) =  0$.  
From now on we will work in this frame. One can easily obtain the corresponding formul\ae~for the density matrix in a generic inertial  frame, $\rho_{\rm gen}$, by inverting the  transformation in~(\ref{rhoRest0}).
  
  \subsection{Ensemble averages} \label{Ensemble averages}
  From~(\ref{rhoRest0}) one obtains the following expressions for the (ensemble) average energy, angular momentum and generic charge: respectively,
  \bea \langle H\rangle &=&  \Tr\left(\rho H\right)=\left.-\partial_\beta \log Z\right|_{\vec\tau, \beta_a}, \label{avHg}\\
   \langle J_k\rangle &=&  \Tr\left(\rho J_k\right)=\left.-\partial_{\tau_k} \log Z\right|_{\beta, \beta_a},\label{avJg} \\
   \langle Q^a \rangle &=& \Tr\left(\rho Q^a\right)=\left.-\partial_{\beta_a} \log Z\right|_{\beta, \vec\tau}.  \label{avQg} \eea
As explicitly indicated,    the derivatives on the right-hand sides are, respectively, taken keeping $\{\vec\tau, \beta_a\}$, $\{\beta, \beta_a\}$ and $\{\beta, \vec\tau\}$ constant. 

Thanks to the symmetry under  rotations we know that the average angular momentum $\langle \vec J\,\rangle = \Tr(\rho\vec J\, )$ is parallel to $\vec\Omega$ (or equivalently $\vec\tau$), namely 
\be \langle\vec J\, \rangle = \alpha_J \vec\Omega, \ee 
where $\alpha_J$ is a function of the rotationally invariant quantities $\beta$, $\mu_a$ and $\Omega\equiv|\vec \Omega|$.

  \section{Free fields}\label{Free fields}

  Let us start by considering a generic number of free fields,  $\varphi_s$ (here all $\varphi_s$ are taken to be real without loss of generality). General interacting theories will be investigated in Sec.~\ref{Interacting theories and general path-integral formula}. The Lagrangian is here given by 
 \be \mathscr{L}= \frac12\partial_\mu \varphi \partial^\mu\varphi - \frac12 \varphi M^2 \varphi, 
  \label{freeLag}\ee
where 
$M^2$ is the scalar squared-mass matrix. 
A vector notation is used, $\varphi$ is a field array with components $\varphi_s$ and the transpose operation is understood.

The group $\mathcal{G}$ acts on $\varphi$ as follows: 
\be \varphi \to \exp(i\alpha_a\theta^a) \varphi \ee
for some real parameters $\alpha_a$, 
where the $\theta^a$ are the generators of $\mathcal{G}$ in the representation of the scalars. Since the $\varphi_s$ are taken to be real, $\theta^a$ are purely imaginary and antisymmetric: this ensures that the group has unitary elements and transforms a real array of scalars into a real array of scalars. The invariance of the mass terms in~(\ref{freeLag}) implies $[\theta^a,M^2]=0$, which in turn tells us (by Schur's Lemma) that $M^2$ can be taken to be block diagonal with each block proportional to the identity matrix;
 the different blocks correspond to irreducible representations of $\mathcal{G}$. 
 This allows us to consider, at least for free fields, the various irreducible representations separately as we do from now on in this Sec.~\ref{Free fields}. All fields belonging to the same irreducible representation have of course the same mass, which in the following is denoted\footnote{The letter $m$ is not used for the mass here because it is used for the angular-momentum quantum number, see below.} $\mu$. The generators of $\mathcal{G}$ in the given irreducible representation are denoted $R^a$.

The corresponding field operator $\Phi$ is the most general solution of the Klein-Gordon equation
\be (\Box +\mu^2)\Phi = 0, \label{KG}\ee 
with $\Box\equiv\partial_\mu\partial^\mu$, satisfying the canonical commutation relations
\be [\Phi_s(t,\vec x), P_{\varphi\, s'}(t,\vec y)] = i\delta_{ss'} \delta(\vec x-\vec y), \quad  [\Phi(t,\vec x), \Phi(t,\vec y)] = 0, \quad  [P_{\varphi}(t,\vec x), P_{\varphi}(t,\vec y)] = 0, \label{CanComm}\ee
where $P_{\varphi} = \dot\Phi$ is the conjugate momentum operator and a dot denotes a time derivative. As well known, Eqs.~(\ref{KG}) and~(\ref{CanComm}) imply that $\Phi$ can be written in terms of creation and annihilation operators. Here, however, it is convenient to use the unusual cylindrical coordinates 
\be x^1 = r\cos\phi, \qquad x^2 = r\sin\phi, \qquad x^3 = z\label{CyCoo} \ee
instead of the Cartesian coordinates, $\{x^1, x^2, x^3\}$, with the $z$ axis chosen to be parallel to $\vec\Omega$. 
In cylindrical coordinates~\cite{Vilenkin:1980zv}
\be \Phi_s(t,\vec x) =\sum_{m=-\infty}^{+\infty}\int_\mu^\infty d\omega\int_{-p_0}^{p_0}dp \left[a_{\omega p m s} \varphi_{\omega p m}(\vec x) e^{-i\omega t} +\mbox{h.c.}\right], \label{PhiDec} \ee
where the $a_{\omega p m s}$ are the annihilation operators of particles with energy $\omega$, linear and angular  momentum along the third axis $p$ and $m$, respectively, and species $s$, 
\be [a_{\omega p m s}, a_{\omega' p' m' s'}^\dagger] = \delta_{ss'} \delta_{mm'}\delta(\omega - \omega')\delta(p-p') \label{Commaad}\ee
and
\be \varphi_{\omega p m}(\vec x) e^{-i\omega t} = \frac{J_m(\alpha r)}{2^{3/2}\pi} e^{ipz+im\phi-i\omega t} \label{WFcyl}\ee 
are the corresponding wave functions, with $\alpha \equiv \sqrt{p_0^2-p^2}$, 
$p_0\equiv \sqrt{\omega^2 -\mu^2}$ and the $J_m$ being the cylindrical Bessel functions. A way to show that~(\ref{PhiDec}) is the  correct expression is to note that a generic linear combination of the wave functions in~(\ref{WFcyl}) is the general regular solution of the Klein-Gordon equation~(\ref{KG}) and the canonical commutators in~(\ref{CanComm}) imply~(\ref{Commaad}) with the normalization given in~(\ref{WFcyl}).

\subsection{Computing ensemble averages} \label{Computing ensemble averages}

In Eqs.~(\ref{avHg}),~(\ref{avJg}) and~(\ref{avQg}) we have presented the general formul\ae~to compute the (ensemble) averages of the observables $H$, $J_k$ and $Q^a$ for arbitrary equilibrium density matrices. In this section we provide a general method to  compute averages in the case of free fields. In more physical terms the case we are going to consider in this section is that of systems with negligibly small interactions. 

To facilitate the computation of averages let us perform a change of basis in the space of particle states as follows. 
Note that $\mathcal{G}$ acts on $\Phi(x)$ as follows:
\be \exp(i\alpha_aQ^a)\Phi(x)\exp(-i\alpha_aQ^a) = \exp(i\alpha_aR^a)\Phi(x), \label{GonPhi} \ee
which corresponds to the following action on the annihilation and creation operators,
\bea  \exp(i\alpha_aQ^a)a_{\omega p m s}\exp(-i\alpha_aQ^a) = \exp(i\alpha_aR^a)_{ss'}a_{\omega p m s'}, \\ 
\exp(i\alpha_aQ^a)a^\dagger_{\omega p m s}\exp(-i\alpha_aQ^a) = \exp(i\alpha_aR^a)_{ss'}a^\dagger_{\omega p m s'} . \label{Gonad} \eea
This implies the following action of $\mathcal{G}$  on one-particle states $|\omega,p,m,s\rangle$, namely states that are obtained by applying $a_{\omega p m s}^\dagger$ on the vacuum  (these are one-particle states with energy $\omega$, linear and angular  momentum along the third axis $p$ and $m$, respectively, and species $s$):
  \be  \exp(i\alpha_aQ^a) |\omega,p,m,s\rangle =  \exp(i\alpha_aR^a)_{ss'} |\omega,p,m,s'\rangle, \label{Gon1P} \ee
  where the invariance of the vacuum under $\mathcal{G}$ was used.
The expression above implies, among other things, 
  \be \mu_a Q^a|\omega,p,m,s\rangle =  (\mu_a R^a)_{ss'} |\omega,p,m,s'\rangle. \ee
    Now, by performing a $\mu_a$-dependent unitary transformation of these states,
  \be |\omega,p,m;d\rangle\equiv U_{ds} |\omega,p,m,s\rangle, \ee 
  with the $U_{ds}$ satisfying $U_{ds}U^*_{d's}= \delta_{dd'}$, it is possible to diagonalize the matrix $\mu_aR^a$:
  \be U\mu_a R^aU^\dagger = \mathcal{M}, \label{fromRtoD} \ee 
  where $U$ is the matrix with elements $U_{ds}$ and $\mathcal{M}$ is a (generically $\mu_a$-dependent) diagonal real matrix. Therefore, in the new   basis 
    \be \mu_a Q^a|\omega,p,m;d\rangle =  \mathcal{M}_d |\omega,p,m; d\rangle, \label{muQM}\ee
    where the $\mathcal{M}_d$ are the diagonal elements of $\mathcal{M}$, which encode the effect of the chemical potentials, and in the right-hand side of~(\ref{muQM}) there is no sum over the index $d$.

  This basis is convenient to compute the partition function and the averages of the product of two annihilation and creation operators and of the observables $H$, $\vec J$ and $Q^a$. To this purpose, another convenient thing to do is to discretize the variables  $\omega$ and $p$  such that integrals over these quantities become sums: this can be done by dividing the ranges of $\omega$ and $p$ in small discrete steps of size $\Delta\omega$ and  $\Delta p$  and then let $\Delta\omega\to0$ and $\Delta p\to0$ to recover the continuum case. Such discretization can be obtained by putting the system in a finite volume, for example, of cylindrical shape (with axis in the direction of $\vec\Omega$) and imposing the periodicity condition $e^{ip(z+L)}=e^{ipz}$ and the continuity condition $J_m(\alpha R)=0$, where $R$ and $L$ are the radius and height of the cylinder. As usual $e^{ip(z+L)}=e^{ipz}$ implies that $p$ can only take values that are integer multiples of $2\pi/L$. To find the discrete values of $\omega$ note that $J_m(\alpha R) = 0$
   implies that $\alpha$ is quantized: for each value of $m$, the quantity $\alpha$  can only assume values 
  \be \alpha_{m,n} \equiv \frac{j_{m,n}}{R}, \label{alphadef}\ee
   where $j_{m,n}$ is the $n$th positive zero of the function $J_m$. So  for each $m$ and $p$ the energy can only assume the discrete values 
\be \omega_{m,n}(p) \equiv \sqrt{\mu^2+ \alpha_{m,n}^2 + p^2},\qquad n = 1,2,3,...\, . \label{omegamn} \ee 
In the discretized version, for any integrand $[...]$ one should actually read
\be \SumInt_q [...]\equiv \sum_{m=-\infty}^{+\infty}\int_\mu^\infty d\omega\int_{-p_0}^{p_0}dp \left[...\right] = \sum_{m=-\infty}^{+\infty}\sum_{j=-\infty}^{+\infty}\frac{2\pi}{L} \sum_{n=1}^\infty  \Delta \omega_{m,n}(p) \left[...\right]^{\omega\to\omega_{m,n}(p)}_{p\to 2\pi j/L} , \label{DicMes} \ee
where 
\be \Delta \omega_{m,n}(p) \equiv \omega_{m,n+1}(p) - \omega_{m,n}(p), \ee 
and we have also introduced a shorthand notation, with $q$ representing the set of variables $\{\omega, p, m\}$. As explicitly indicated in the right-hand side of~(\ref{DicMes}), the continuous variables $\omega$ and $p$ should be substituted by the respective discretized versions, $\omega_{m,n}(p)$ and $2\pi j/L$,   in the generic integrand  $[...]$.
  
  One can now introduce the rescaled annihilation operators $\alpha_{\omega p m, d} \equiv \sqrt{\Delta\omega\Delta p} \, a_{\omega p m, d}$ (with $a_{\omega p m, d}\equiv U_{ds}^*a_{\omega p m s}$), which satisfy
    \be [\alpha_{\omega p m, d}, \alpha_{\omega' p' m', d'}^\dagger] = \delta_{dd'}\delta_{mm'} \delta_{\omega \omega'}\delta_{pp'}. \label{CommaadD}\ee

    The density matrix in~(\ref{rhoRest}) can then be expressed in terms of the number operators 
    \be N_{\omega p m d} \equiv \alpha^\dagger_{\omega p m, d}\alpha_{\omega p m, d}\ee 
    as follows:
    \be \rho=\frac1{Z} \exp\left(-\beta \sum_{\omega p m d}\left( \omega  -m\Omega   - \mathcal{M}_d  \right)N_{\omega p m d}\right) . \label{rhofre} \ee 
    The  $\mathcal{M}_d$  represent the contribution of the chemical potentials $\mu_a$. The quantity in~(\ref{rhofre}) is nothing but the density matrix with zero thermal vorticity and chemical potentials, but with energies $\omega$ replaced by $\omega - m\Omega - \mathcal{M}_d$. As a result the partition function is 
    \be Z = \prod_{\omega p m d} \frac1{1-e^{-\beta\left(\omega-m\Omega - \mathcal{M}_d\right)}}. \label{Zpart}\ee
    Then, using  
\be \log Z =  - \sum_{\omega p m d} \log \left(1- e^{-\beta(\omega - m\Omega   - \mathcal{M}_d)}  \right), \label{logZ}\ee 
one finds
\bea \langle \alpha_{\omega p m, d}^\dagger \alpha_{\omega' p' m', d'}\rangle &=& f_B(\omega - m\Omega - \mathcal{M}_d)\delta_{dd'} \delta_{mm'} \delta_{\omega \omega'}\delta_{pp'}, \label{dalpha}\\ 
\langle \alpha_{\omega p m, d} \alpha^\dagger_{\omega' p' m', d'}\rangle &=& (1+f_B(\omega - m\Omega - \mathcal{M}_d)) \delta_{dd'} \delta_{mm'}\delta_{\omega \omega'}\delta_{pp'}, \label{alphad}\eea
where
\be f_B(x) \equiv \frac1{e^{\beta x}-1} \label{fBdef}\ee
is the Bose-Einstein distribution. Setting $d=d'$, $m=m'$, $\omega=\omega'$ and $p=p'$ in~(\ref{dalpha}) one finds the average numbers of particles. Moreover,  as usual, the average of the product of two annihilation or two creation operators vanish,
\be \quad \langle \alpha_{\omega p m, d} \alpha_{\omega' p' m', d'}\rangle = \langle \alpha_{\omega p m, d}^\dagger \alpha_{\omega' p' m', d'}^\dagger\rangle = 0. \ee
Let us recall that in~(\ref{Zpart}) a single irreducible representation of $\mathcal{G}$ is considered: to obtain the partition function for all  irreducible representations one can simply take the product of all partition functions of single irreducible representations.

Moreover, using~(\ref{logZ}) and the general expressions~(\ref{avHg}),~(\ref{avJg}) and~(\ref{avQg}), the average values of $H$, $\vec J$  and $Q^a$ turn out to be 
\bea &&\langle H\rangle =  
\sum_{\omega p m d} \omega f_B(\omega - m\Omega - \mathcal{M}_d), \quad   \langle  J_z\rangle =  
\sum_{\omega p m d} m f_B(\omega - m\Omega - \mathcal{M}_d), \label{HJz}\\
&&\langle Q^a\rangle =  
\sum_{\omega p m d} \frac{\partial\mathcal{M}_d}{\partial\mu_a}  f_B(\omega - m\Omega - \mathcal{M}_d) \label{avQ}.\eea
In the limit $\Delta p\to0$ (which corresponds to $L\to\infty$) these expressions allow us to compute  the average values of the energy, angular momentum and charges per unit of length in the $z$ direction in terms of integrals rather than sums over $p$.
The limit $\Delta \omega\to0$ (which corresponds to $R\to\infty$) is more delicate and will be discussed below.

Before doing so let us note that we can also go back to the original basis by inverting~(\ref{fromRtoD}) and obtain, starting from~(\ref{dalpha})-(\ref{alphad}), 
  \bea \langle \alpha_{\omega p m s}^\dagger \alpha_{\omega' p' m' s'}\rangle &=& f_B(\omega - m\Omega - \mu_a R^a)_{ss'}\delta_{mm'} \delta_{\omega \omega'}\delta_{pp'},\label{Avada} \\ 
\langle \alpha_{\omega p m s} \alpha^\dagger_{\omega' p' m' s'}\rangle &=& (1+f_B(\omega - m\Omega - \mu_a R^a))_{s's}\delta_{mm'} \delta_{\omega \omega'}\delta_{pp'}, \label{Avaad}\eea
with $\alpha_{\omega p m s} \equiv U_{ds}\alpha_{\omega p m, d}$, and, of course,
\be \quad \langle \alpha_{\omega p m s} \alpha_{\omega' p' m' s'}\rangle = \langle \alpha_{\omega p m s}^\dagger \alpha_{\omega' p' m' s'}^\dagger\rangle = 0. \label{Avaa}\ee
The quantity $f_B(\omega - m\Omega - \mu_a R^a)$ appearing in the expressions above is the function $f_B$ in~(\ref{fBdef}) computed in the matrix $\omega-m\Omega- \mu_a R^a$ (where an identity matrix multiplying $\omega-m\Omega$ is understood). In practice, to compute this quantity one can determine the projectors $\mathcal{P}_d$ associated with the eigenvalues $\mathcal{M}_d$ of  $\mu_a R^a$ and use the well-known spectral decomposition
\be f_B(\omega - m\Omega - \mu_a R^a) = \sum_d f_B(\omega - m\Omega - \mathcal{M}_d)\mathcal{P}_d,\ee 
where the projectors always satisfy $\sum_d\mathcal{P}_d =1$, $\mathcal{P}_d\mathcal{P}_{d'} = \delta_{dd'}\mathcal{P}_{d'}$ and $\mu_a R^a = \sum_d\mathcal{M}_d \mathcal{P}_d$. 

To have a feeling of how the $\mathcal{M}_d$ and $\mathcal{P}_d$ may look like in some non-Abelian case, consider $\mathcal{G} =$ SO(3) and a triplet scalar representation, so
\be R^1=i  
\left(\begin{array}{ccc}
 0 & 0 & 0 \\
 0 & 0 & -1 \\
 0 & 1 & 0 \\
\end{array}\right), \quad  R^2=i \left(
\begin{array}{ccc}
 0 & 0 & 1 \\
 0 & 0 & 0 \\
 -1 & 0 & 0 \\
\end{array}
\right), \quad R^3=i \left(
\begin{array}{ccc}
 0 & -1 & 0 \\
 1 & 0 & 0 \\
 0 & 0 & 0 \\
\end{array}
\right).
\ee
In this case the eigenvalues of $\mu_aR^a$ are $\mathcal{M}_0 =0$ and $\mathcal{M}_{\pm} = \pm \sqrt{\mu_a\mu_a}$ and the corresponding projectors read
\be \mathcal{P}_0=\left(
\begin{array}{ccc}
 \frac{\mu_1^2}{\mu_a\mu_a}& \frac{\mu_1 \mu_2}{\mu_a\mu_a} & \frac{\mu_1 \mu_3}{\mu_a\mu_a} \\
 \frac{\mu_1 \mu_2}{\mu_a\mu_a} & \frac{\mu_2^2}{\mu_a\mu_a} & \frac{\mu_2 \mu_3}{\mu_a\mu_a} \\
 \frac{\mu_1 \mu_3}{\mu_a\mu_a} & \frac{\mu_2 \mu_3}{\mu_a\mu_a} & \frac{\mu_3^2}{\mu_a\mu_a} \\
\end{array}
\right), \quad \mathcal{P}_\pm=\left(
\begin{array}{ccc}
 \frac{\mu_2^2+\mu_3^2}{2 \mu_a\mu_a} & -\frac{\mu_1 \mu_2+i\mu_3\mathcal{M}_\pm }{2 \mu_a\mu_a} & \frac{i\mu_2\mathcal{M}_\pm-\mu_1 \mu_3}{2 \mu_a\mu_a} \\
 \frac{i\mu_3\mathcal{M}_\pm-\mu_1 \mu_2}{2 \mu_a\mu_a} & \frac{\mu_1^2+\mu_3^2}{2 \mu_a\mu_a} & -\frac{i\mu_1\mathcal{M}_\pm +\mu_2 \mu_3}{2 \mu_a\mu_a} \\
 -\frac{i\mu_2\mathcal{M}_\pm+\mu_1 \mu_3}{2 \mu_a\mu_a} & \frac{i\mu_1\mathcal{M}_\pm-\mu_2 \mu_3}{2 \mu_a\mu_a} & \frac{\mu_1^2+\mu_2^2}{2 \mu_a\mu_a} \\
\end{array}
\right).\ee

The averages in the left-hand sides of~(\ref{Avada}) and~(\ref{Avaad})  were only expressed in terms of series in~\cite{Becattini:2020qol}. Here, a useful closed form is found. 

It is important to note that the convergence of the averages requires $\Omega$ to be less than a critical value. To understand this point let us consider $\log Z$ (given in~(\ref{logZ})), which can be used to extract the averages as explained above. 
If one separately considers the sum over $m$, the convergence of the series at $m\to +\infty$ implies 
\be \Omega <  \lim_{m\to +\infty}\frac1{m}\min_{\omega\, d} ( \omega - \mathcal{M}_d). \ee 
Using now the expression for the energies in~(\ref{omegamn}), this condition can be written as follows:
\be \Omega <  \lim_{m\to +\infty}\frac1{m}\min_{n\, p\, d} ( \omega_{m,n}(p) - \mathcal{M}_d) = \lim_{m\to +\infty} \frac{j_{m,1}}{m R} = \frac1{R},  \label{ConvCond} \ee 
where we have used that $j_{m,1}/m$ converges to 1 as $m\to+\infty$.
From~(\ref{HJz}) and~(\ref{avQ}) one can clearly see that this condition is implied by the physical requirement that the average energy, angular momentum and charges should be finite.
The large-$R$ limit should, therefore, be taken together with the small-$\Omega$ limit, but, as will be shown shortly, we can obtain a non-vanishing value of the average angular momentum along the $z$ axis by keeping in the limit a non-vanishing and finite value of the  (rotational) velocity parameter 
\be v\equiv \Omega R \ee
 in the interval $0\leq v<1$.

%
%
%

\begin{figure}[t]
\begin{center}
  \includegraphics[scale=0.65]{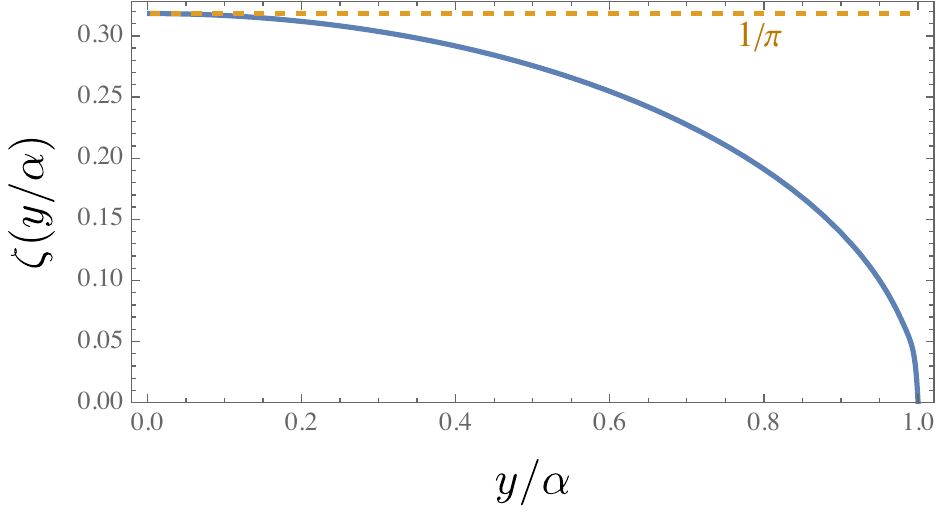}
    \caption{\em The even measure function $\zeta(y/\alpha)$ that determines the momentum space integral. 
    }\label{zeta}
  \end{center}
\end{figure}

In this limit one can think  of $m/R$ as a continuous variable that takes non-vanishing values only for infinite $m$, for which $\alpha_{m,n}\geq \alpha_{m,1}= |m|/R$, where the modulus appears because $j_{m,n}=j_{-m,n}$. So  we can substitute $m/R$ with $y$, where $y\in[-\alpha,\alpha]$. The $y$-independent  variation of $\alpha$  is the difference 
\be \Delta\alpha_{m,n}\equiv \frac{j_{m,n+1}-j_{m,n}}{R}. \label{Deltaalpha}\ee
 For large $n$ and fixed $m$ (which corresponds to $|y|\ll\alpha$) the variable $\alpha$ (see~(\ref{alphadef})) acquires non-vanishing values even in the large-$R$ limit  and, using McMahon's  Asymptotic Expansions of the $j_{m,n}$ for large $n$, one finds $R\Delta\alpha_{m,n}\to \pi$. 
 On the other hand, using asymptotic expansions of the $j_{m,n}$ for large $m$ and fixed $n$ (which corresponds to $|y|\simeq\alpha$) one finds that  $R\Delta\alpha_{m,n}$ grows indefinitely in this limit and so its inverse tends to zero.
  For arbitrary values of $y$, $R\Delta\alpha_{m,n}$ goes to an even function of the dimensionless ratio $y/\alpha$, which we call $\zeta^{-1}(y/\alpha)$ and can be determined numerically (see Fig.~\ref{zeta}). A table containing the numerical determination of $\zeta$ as a function of $y/\alpha$ can be found at~\cite{dataset}.

Thus, using the first expression in~(\ref{HJz}), the average of the energy density $\rho_E$ (energy per unit of space volume) is
\be \langle \rho_E\rangle = \sum_d\int \frac{\zeta(y/\alpha)d\alpha dy dp}{2\pi^2} \,  \omega  f_B (\omega -  v y- \mathcal{M}_d), \label{rhoE}\ee
where  $\omega = \sqrt{\mu^2+\alpha^2 +p^2}$ and  the integral is over the full momentum space, $\alpha\in[0,\infty), y\in[-\alpha,\alpha], p\in (-\infty,\infty)$.  It is convenient to change integration variables defining $\xi\equiv y/\alpha\in[-1,1]$ so that  
\be \langle \rho_E\rangle = \sum_d\int \frac{\alpha\zeta(\xi) d\alpha d\xi dp}{2\pi^2} \,  \omega  f_B (\omega -  v \alpha \xi- \mathcal{M}_d). \label{rhoEp}\ee
We checked that when $v=0$ the well-known expression of $\langle \rho_E\rangle$ in the absence of average angular momentum is reproduced by integrating over $\xi\in[-1,1]$.  Analogously, using the second expression in~(\ref{HJz}), the average of the angular momentum density per  unit of distance from the
 rotation axis, $\mathcal{J}_z$, is given by
\be \langle \mathcal{J}_z\rangle =   \sum_d\int \frac{\alpha \zeta(\xi)d\alpha d\xi dp}{2\pi^2} \,  \alpha \xi f_B (\omega  - v \alpha \xi - \mathcal{M}_d) \label{CallJz}\ee
and, using~(\ref{avQ}), the averages of the charge densities (the charges per unit of space volume) are 
\be \langle\rho_a\rangle =   \sum_d\frac{\partial\mathcal{M}_d}{\partial\mu_a}\int \frac{\alpha\zeta(\xi)d\alpha d\xi dp}{2\pi^2} \,  f_B (\omega  -v \alpha\xi - \mathcal{M}_d). \label{avQcon}\ee

The convergence of the averages also imply that the quantity in the argument of $f_B$ in~(\ref{rhoE}) must never vanish (and so must be positive) in the full integration domain. This results in the condition 
\be \mathcal{M}_d <  \mu \sqrt{1-v^2}, \qquad (\mbox{for all}~d)\label{BoundChem}\ee
which is a generalization of the known result for $v=0$ and a single chemical potential that the chemical
potential has to be less than the lowest single-particle energy in the bosonic case. 
Note that for $v\neq 0$ the condition becomes stronger.

One obtains the full energy and $a$-th charge of the system by integrating, respectively, the averages $\langle \rho_E\rangle$ and $\langle\rho_a\rangle$ over the space region of the plasma. The full angular momentum is instead obtained by integrating the average $\langle\mathcal{J}_z\rangle$ multiplied by the distance from the rotation axis over the space region of the plasma.

\begin{figure}[t]
\begin{center}
  \includegraphics[scale=0.5]{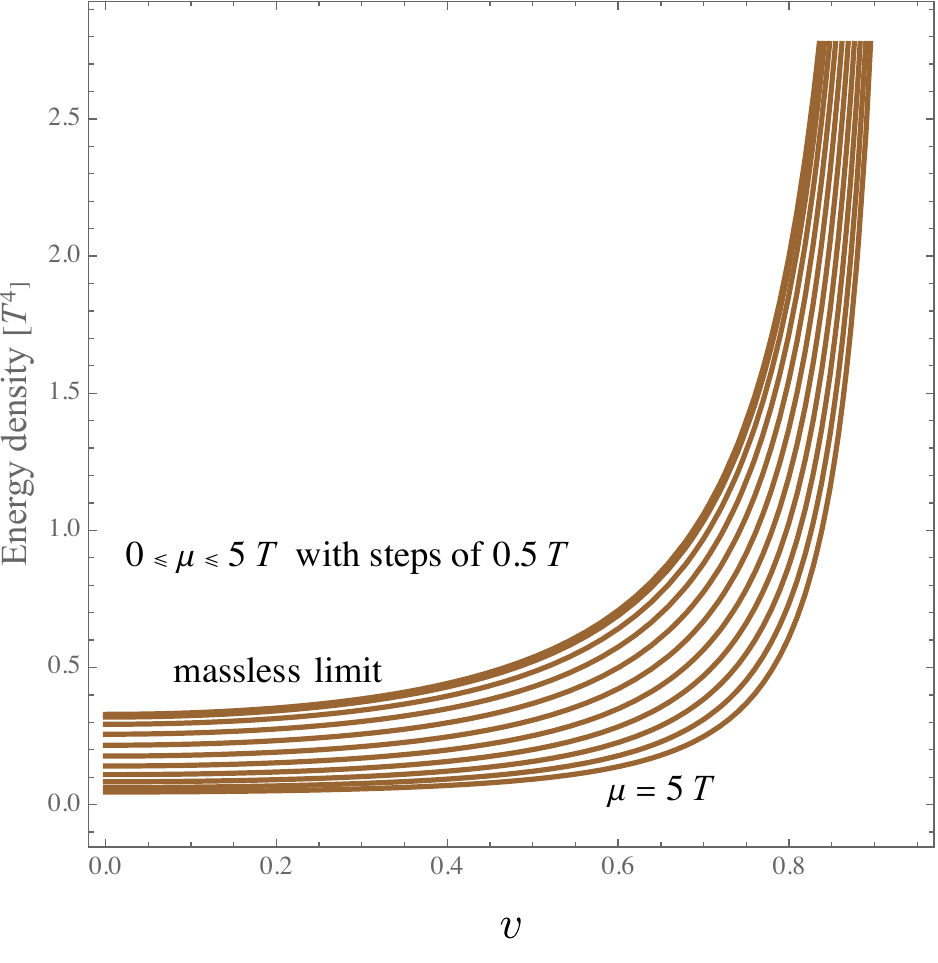}  \hspace{1cm} \includegraphics[scale=0.5]{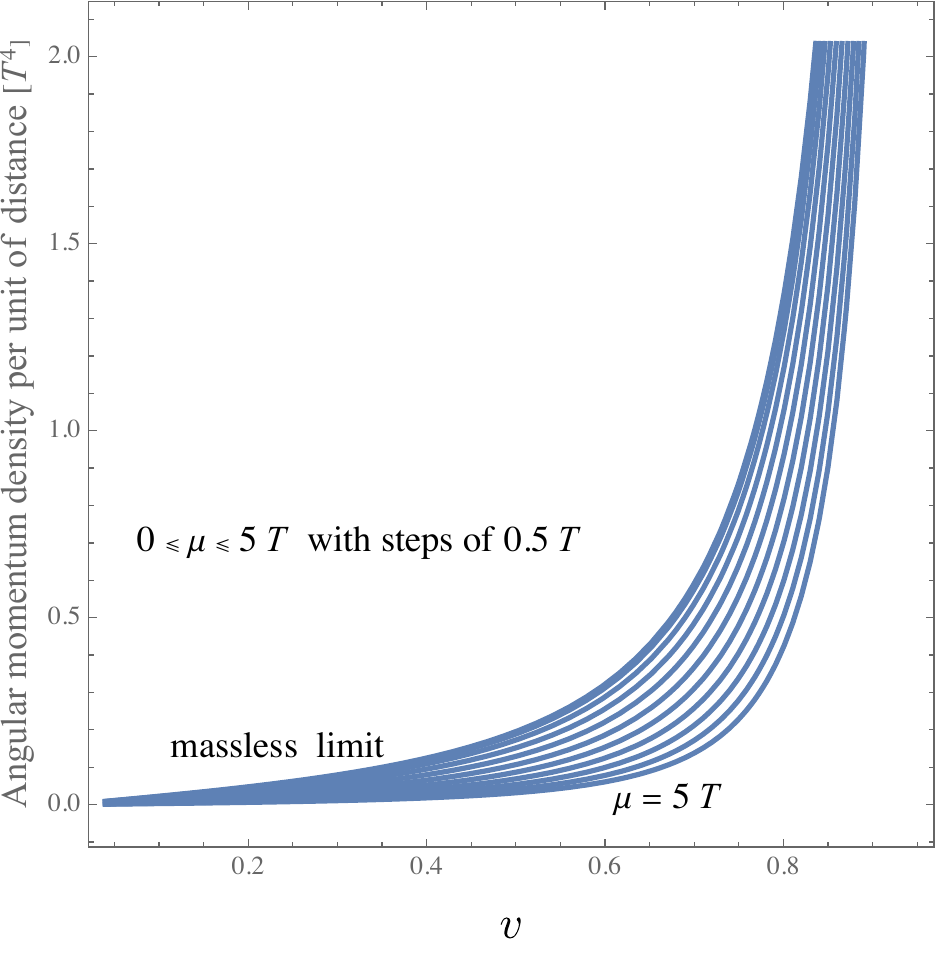}  
    \caption{\em Average energy density (left plot) and average angular momentum density per unit of distance from the rotation axis (right plot) as a function of the (rotational) velocity parameter $v$  for several values of the mass $\mu$. A single real scalar field is considered.}\label{rhoEJ}
  \end{center}
\end{figure}

In Fig.~\ref{rhoEJ} the quantities $\langle \rho_E\rangle$ and $\langle \mathcal{J}_z\rangle$ for a single real scalar are shown as functions of $v$ for several values of $\mu$. One can obtain all positive values of $\langle \mathcal{J}_z\rangle$ by varying $v$ in $[0,1)$. Then $\langle \rho_E\rangle$ is predicted. Note that both $\langle \rho_E\rangle$ and $\langle \mathcal{J}_z\rangle$  increase indefinitely as 
$v$ approaches $1$. Tables containing the numerical determination of $\langle \rho_E\rangle$ and $\langle \mathcal{J}_z\rangle$ as a function of $v$ for the case considered in Fig.~\ref{rhoEJ} can be found at~\cite{dataset}. Particles in the coronas of black holes can orbit at velocities approaching the speed of light, $v_{\rm max}\simeq 1$. 
For these systems the inclusion of the effects of $v$ is thus essential.

\subsection{Thermal propagator}\label{Thermal propagator}

In TFT one is interested in computing thermal Green's functions. These are indeed very useful tools to compute the rates of inclusive processes (such as decay or scattering rates
summed over final states or production rates summed over initial states)~\cite{Salvio:2024upo}.

General Green's functions will be studied in Sec.~\ref{Interacting theories and general path-integral formula} including arbitrary interactions. Here the simplest Green's function (the thermal propagator) is studied for free fields as it plays a crucial role in perturbation theory. This function is defined by 
\bea \langle {\cal T}\Phi_s(x_1)\Phi_{s'}(x_2)\rangle&\equiv&\theta(t_1-t_2)\langle \Phi_s(x_1)\Phi_{s'}(x_2)\rangle+\theta(t_2-t_1)\langle \Phi_{s'}(x_2)\Phi_s(x_1)\rangle, \label{ThermalProp}
 \eea
 where $t_{1,2}$ are the time components of the spacetime points $x_{1,2}$ and $\theta$ is the Heaviside step function. 
 
Using the averages of the product of pairs of annihilation and creation operators determined in Sec.~\ref{Computing ensemble averages} one obtains, for the  ``non time-ordered" 2-point function of a field $\Phi$ in an arbitrary irreducible representation of $\mathcal{G}$,
\bea G_{ss'}^>(x_1,x_2)&\equiv& \langle \Phi_s(x_1)\Phi_{s'}(x_2)\rangle \nonumber \\ &=& \SumInt_q~\,\left[ (1+f_B(\omega - m\Omega - \mu_a R^a))_{s's}\varphi_q(\vec x_1) \varphi_q^*(\vec x_2)e^{-i\omega (t_1-t_2)} \right.\nonumber \\ 
&& \hspace{2cm}\left.  + f_B(\omega - m\Omega - \mu_a R^a)_{ss'} \varphi_q^*(\vec x_1) \varphi_q(\vec x_2)e^{i\omega (t_1-t_2)} \right],  \label{Wf} \eea
where $\varphi_q$ is a shorthand notation for $\varphi_{\omega p m}$.
We recall that $R^a$ are the generators of $\mathcal{G}$ in the generic irreducible representation under study and $\mu$ the corresponding mass.  The other ``non time-ordered" 2-point function appearing in the thermal propagator in~(\ref{ThermalProp}), i.e.
\be G_{ss'}^<(x_1,x_2)\equiv \langle \Phi_{s'}(x_2)\Phi_{s}(x_1)\rangle \ee 
can  be easily extracted from $G_{ss'}^>(x_1,x_2)$ using the property $G_{ss'}^<(x_1,x_2) = G_{s's}^>(x_2,x_1)$.

In~\cite{Vilenkin:1980zv} a different 2-point function was computed: the thermal average of the time-ordered product (at different spacetime points) of the Matsubara field operator for vanishing $\mu_a$. This does not coincide with $\Phi_s$ but with the field $\tilde \Phi_s$, which will be defined in~(\ref{tildePhi}), after the substitution  $it\to \tau$, where $\tau$ is the imaginary time. The expression obtained here is more general because includes generic values of the chemical potentials and is also more relevant as the generator of time translations is $H$ rather than $H-\vec\Omega\cdot\vec J$.

    \section{General interacting theories and path-integral formula}\label{Interacting theories and general path-integral formula}
  
Let us consider a theory with a generic number of scalar fields, which we collectively represent by an array\footnote{In order to write more compact expressions, the index of $\Phi$, which the internal symmetry group can act on, is understood. The field $\Phi$ can include here an arbitrary number of irreducible representations of the group $\mathcal{G}$.} $\Phi$ of Hermitian field operators. As already mentioned in Sec.~\ref{Thermal propagator}, we are interested~\cite{Salvio:2024upo} in computing thermal Green's functions
\bea \langle {\cal T}\Phi(x_1)...\Phi(x_n)\rangle = \Tr(\rho {\cal T}\Phi(x_1)...\Phi(x_n)),  \label{GreenFunc}
 \eea
 where ${\cal T}$ is the time-ordered product. In Sec.~\ref{Thermal propagator} the case of the thermal propagator ($n=2$) for free fields was considered. Here we investigate the general interacting case for an arbitrary  $n$-point Green's function by using path-integral methods. 
 
 Before proceeding, let us recall that $\Phi(x)$ is the field operator in the  Heisenberg picture, $\Phi(t,\vec x)\equiv \exp(iHt)\Phi(0,\vec{x})\exp(-iHt)$. Being Hermitian, $\Phi(0,\vec{x})$ admits a complete set of eigenstates $|\varphi\rangle$, namely $\Phi(0,\vec{x})|\varphi\rangle = \varphi(0,\vec{x}) |\varphi\rangle$, where $\varphi(0,\vec{x})$ is the corresponding eigenvalue. 
 
 In order to derive a path-integral expression for~(\ref{GreenFunc}), instead of using only $\Phi(t,\vec x)$ and the corresponding eigenstates $|\varphi,t\rangle\equiv \exp(iHt)|\varphi\rangle$, it is  convenient to introduce
 \be \tilde \Phi(t,\vec x) \equiv  \exp(i(H-\vec\Omega\cdot \vec J\,)t)\,\Phi(0,\vec{x})\,\exp(-i(H-\vec\Omega\cdot \vec J\,)t) = \Phi(t, R^{-1}(t\vec\Omega)\vec x) \label{tildePhi}\ee
 and the corresponding eigenstates
  \be |\tilde\varphi,t\rangle \equiv \exp(i(H-\vec\Omega\cdot \vec J\,)t) |\varphi\rangle, \label{tildevar}\ee
 which contain both $H$ and $\vec\Omega$. Note that $|\tilde\varphi,t\rangle$ is an eigenstate of $\tilde \Phi(t,\vec x)$ with eigenvalue $\varphi(0,\vec x)$. The quantity $R(t\vec\Omega)$ in~(\ref{tildePhi}) is the $3\times3$ orthogonal matrix with unit determinant representing the rotation associated with $t\vec\Omega$. Inverting~(\ref{tildePhi}) leads to\footnote{The inversion can be performed as follows. Consider first
 \be \exp(-i\vec\Omega\cdot \vec J\, t')\,\tilde\Phi(t,\vec{x})\,\exp(i\vec\Omega\cdot \vec J\,t') = \Phi(t, R^{-1}((t+t')\vec\Omega)\vec x),\ee
 where $t'$ is an arbitrary time. On the other hand, one also has
 \be \exp(-i\vec\Omega\cdot \vec J\, t')\,\tilde\Phi(t,\vec{x})\,\exp(i\vec\Omega\cdot \vec J\,t')  = \tilde\Phi(t, R^{-1}(t'\vec\Omega)\vec x), \ee
 so, setting $t'=-t$, one finds~(\ref{xomega}).}
 \be \Phi(x) = \tilde\Phi(x^\omega), \qquad \mbox{where} \quad x^\omega\equiv \{t, R(t\vec\Omega)\vec x\}.\label{xomega}\ee
   
   We choose a standard normalization of the states in~(\ref{tildevar}) such that 
\be \int \delta\varphi(t) |\tilde\varphi,t\rangle\langle\tilde\varphi,t| = 1, \qquad \mbox{where} \qquad \delta\varphi(t) \equiv \prod_{\vec{x}} d\varphi(t,\vec{x}).\ee
Under $\mathcal{G}$ the field operator $\Phi(x)$ transforms as follows
\be \exp(i\alpha_aQ^a)\Phi(x)\exp(-i\alpha_aQ^a) = \exp(i\alpha_a\theta^a)\Phi(x), \label{GonPhip} \ee
 and the same is true for $\tilde \Phi(x)$ because $\mu_a Q^a$ commutes with $H$ and $\vec \Omega \cdot \vec J$. Let us denote 
\be \exp(-i\alpha_aQ^a)|\tilde \varphi,t\rangle \equiv |e^{i\alpha_a\theta^a}\tilde\varphi,t\rangle \ee
because this state is an eigenstate of $\tilde\Phi(x)$ with eigenvalue $\exp(i\alpha_a\theta^a) \varphi(0,\vec x)$. So  
\be \langle\tilde \varphi,t| \exp(i\alpha_aQ^a)= \langle e^{i\alpha_a\theta^a}\tilde\varphi,t|. \ee
 Choosing now the basis $\{|\tilde\varphi,t_0\rangle\}$ to represent the trace in~(\ref{GreenFunc}), where $t_0$ is an arbitrary time, one finds
\bea \langle {\cal T}\Phi(x_1)...\Phi(x_n) \rangle &=& \int \delta \varphi(t_0) \langle\tilde\varphi,t_0|\rho\,  {\cal T}\Phi(x_1)...\Phi(x_n)|\tilde\varphi,t_0\rangle \nonumber \\
&=&\frac1{Z}\int \delta \varphi(t_0) \langle\tilde\varphi,t_0|e^{-\beta (H-\vec\Omega  \cdot \vec J - \mu_a Q^a)}  {\cal T}\Phi(x_1)...\Phi(x_n)|\tilde\varphi,t_0\rangle \nonumber\\&=&\frac1{Z}\int \delta \varphi(t_0) \langle e^{\beta \mu_a \theta^a} \tilde\varphi,t_0-i\beta|  {\cal T}\Phi(x_1)...\Phi(x_n)|\tilde\varphi,t_0\rangle\nonumber\\&=&\frac1{Z}\int \delta \varphi(t_0) \langle e^{\beta \mu_a \theta^a} \tilde\varphi,t_0-i\beta|  {\cal T}\tilde\Phi(x^\omega_1)...\tilde\Phi(x^\omega_n)|\tilde\varphi,t_0\rangle. \label{ScalarTGF}\eea

The thermal Green's functions have been reduced to the corresponding expressions in the absence of rotation, but with Hamiltonian $H-\vec\Omega\cdot \vec J$ and with spacetime points $x_i^\omega$ (defined in~(\ref{xomega})) rather than $x_i$. 

Using then the well-known results in the absence of rotation (see e.g.~Ref.~\cite{Landsman:1986uw} and references therein), one can now show 
   \be \hspace{-0.7cm}\boxed{\langle {\cal T}\Phi(x_1)...\Phi(x_n) \rangle =\frac1{Z}\int \delta\varphi \, \delta p_\varphi  \, \varphi(x_1^\omega)...\varphi(x_n^\omega) \exp\left(i \int_C d^4x \left(\dot\varphi(x)p_\varphi(x) - \mathcal{H}^\omega_c(\varphi(x), p_\varphi(x))\right)\right),} \label{PIscalars}\hspace{-0.3cm}\ee 
 where a dot represents a time derivative, $p_\varphi$ is the momentum conjugate to $\varphi$ and the path integration is on all field configurations satisfying  the twisted periodicity condition:
  \be e^{\beta\mu_a \theta^a}\varphi(t_0,\vec{x}) = \varphi(t_0-i\beta,\vec{x}).\label{PerCon}\ee
The integration measures on $\varphi$ and $p_\varphi$ are
  \be \delta\varphi = \prod_x d\varphi(x), \quad \delta p_\varphi= \prod_x d(p_\varphi(x)/(2\pi)).\ee
 The function $\mathcal{H}^\omega_c$ is, in the classical limit, the full classical Hamiltonian density, including the effect of rotation. The defining property of this quantity is
  \be \int d^3x \, \mathcal{H}^\omega_c = H_c- \vec\Omega\cdot \vec J_c, \label{callHc}\ee  
  where $H_c$ and $\vec J_c$ are 
\be H_c(\varphi, p_\varphi)\equiv \frac{\langle\varphi|H|p_\varphi\rangle}{\langle\varphi|p_\varphi\rangle}, \qquad \vec J_c(\varphi, p_\varphi)\equiv \frac{\langle\varphi|\vec J |p_\varphi\rangle}{\langle\varphi|p_\varphi\rangle},  \label{HJc}  \ee 
where $|p_\varphi\rangle$ is the generic eigenstate of the conjugate momentum operator $P_{\varphi}$.     
Moreover,  in the path integral, while the space integral has  no restriction,
 the integral over $t$ is performed on a contour $C$ in the complex $t$ plane that connects  $t_0$ and $t_0-i\beta$ and contains the time components $x_1^0, ... , x_n^0$ of   $x_1, ... , x_n$. This is because we have to include $t_0$, $x_1^0, ... , x_n^0$ and $t_0-i\beta$ in the set of discrete times (that we introduce in deriving the path integral). Note that the field insertion $\varphi(x_1^\omega)...\varphi(x_n^\omega)$ in~(\ref{PIscalars}) involves the points $x_i^\omega$.

 The partition function that appears in the denominator in~(\ref{PIscalars}) is
  \be \boxed{Z =\int \delta\varphi \, \delta p_\varphi  \,  \exp\left(i \int_C d^4x \left(\dot\varphi(x)p_\varphi(x) - \mathcal{H}_c^\omega(\varphi(x), p_\varphi(x))\right)\right),} \label{GenPart}  \ee
  which is nothing but the path integral in~(\ref{PIscalars}) without $\varphi(x_1^\omega)...\varphi(x_n^\omega)$. Recall that $Z$ can be used to compute the averages of observables, see e.g.~(\ref{avHg}),~(\ref{avJg}) and~(\ref{avQg}).

  The time-ordered product $\mathcal{T}$ should be understood as the product  ordered according to the orientation of $C$: one introduces a parametrization $t(u)$ of $C$, where $u$ is a real parameter such that $t(u)$ proceeds along $C$ with its orientation as $u$ increases. The ordering along the contour is the ordering in $u$. For example, given two points $t$ and $t'$ on $C$, which correspond respectively to $u$ and $u'$, namely $t=t(u)$ and $t'=t(u')$, the Heaviside step function is defined as
\be \theta(t'-t)\equiv \theta(u'-u). \label{contourT}\ee
 The Dirac delta function $\delta$  on the contour $C$ is defined as follows: given two points $t$ and $t'$ on $C$, which correspond respectively to $u$ and $u'$, 
\be \delta(t-t') \equiv \left(\frac{d t}{d u}\right)^{-1} \delta(u-u'). \label{contourDF}\ee
With this definition we have, for a generic function $f$ on $C$,
\be \int_C dt' \delta(t-t') f(t') = f(t) \ee
and 
\be \frac{d}{dt'}\theta(t'-t) = \delta(t'-t). \ee

The contour $C$ can be chosen just like in the absence of $\vec\Omega$ and one can recover the imaginary-time or real-time formalism with specific choices. The former is obtained by setting $t_0=0$ and going from $0$ to $-i\beta$ along the imaginary axis~\cite{Matsubara:1955ws}. The latter can be obtained, for example, with the choice of~\cite{Matsumoto:1982ry} (see also Fig.~1 of~\cite{Salvio:2024upo} for a picture in the present notation) and implies a doubling of the degrees of freedom.

  As clear from~(\ref{PIscalars}) and~(\ref{callHc}), in the presence of a non-vanishing average angular momentum one has to substitute  
  $H_c$ with $H_c -\vec \Omega\cdot \vec J_c$ in the path integral. It is interesting to note that this substitution is nothing but the transformation rule of the classical Hamiltonian from an inertial frame to a frame rotating with angular-velocity vector $\vec \Omega$.  One can, therefore, identify $\vec \Omega$ with the angular-velocity vector of the rotating plasma.

 Whatever choice we make for $C$ we can generate the thermal Green's function by performing functional derivatives of the generating functional 
 \be \boxed{\mathcal{Z} (j) =\frac1{Z} \int \delta\varphi \, \delta p_\varphi  \, \exp\left(i \int_C d^4x \left(\dot\varphi(x)p_\varphi(x) - \mathcal{H}^\omega_c(\varphi(x), p_\varphi(x))+j(x)\varphi(x^\omega)\right)\right).} \label{GenFun}\ee  
 The explicit formula is
 \be \langle {\cal T}\Phi(x_1)...\Phi(x_n) \rangle  = \left.\frac1{ i^n}\frac{\delta^n}{\delta j(x_1) ... \delta j(x_n)}  \mathcal{Z}(j) \right|_{j=0}, \label{ZJ0}\ee
 where here we use the Dirac delta function  on the contour $C$, Eq.~(\ref{contourDF}), to define functional derivatives.

   \section{Lagrangian path integral for ordinary scalars }\label{MomInt}
  The result in~(\ref{GenFun}) applies to all scalar theories in the presence of generic values of the temperature, thermal vorticity and chemical potentials. 
 
 Let us now consider ordinary scalars for which 
 \be \mathcal{H}_c = \frac{p_\varphi^2}2 +\frac{(\vec\nabla\varphi)^2}{2}+
 \frac{1}2 
\varphi M^2\varphi+\mathcal{V}(\varphi),  \ee
  where  $\mathcal{V}$ is the potential density minus the mass term $ 
\varphi M^2\varphi/2$. This $\mathcal{H}_c$ can be obtained from the quantum Hamiltonian
  \be H = \int d^3x \left(\frac{P_\varphi^2}2 +\frac{(\vec\nabla\Phi)^2}{2}+ \frac{1}2 
\Phi M^2\Phi+\mathcal{V}(\Phi)\right),\label{Hordinary}\ee
  with $P_\varphi$ being the conjugate-momentum field operator, $P_\varphi|p_\varphi\rangle=p_\varphi|p_\varphi\rangle$. 
  
 To proceed we need the explicit expressions of the angular and linear momentum operators; one can take
   \be J_k = \epsilon_{ijk}\int d^3x \, (x^j\partial_i\Phi) P_\varphi, ~ \iff ~\vec J= -\int d^3x \, (\vec x \times \vec\nabla\Phi ) P_\varphi , \qquad  P^i = -\int d^3x \, \partial_i\Phi \, P_\varphi.\ee
  Indeed, these expressions lead to the well-known commutators of $H$, $P^i$ and $J_k$. Using the second definition in~(\ref{HJc}) one finds
  \be \vec J_c = -\int d^3x \, (\vec x \times \vec\nabla\varphi ) p_\varphi. \ee

Let us now perform the integration over the momenta.
  In the present case $\mathcal{H}_c^\omega$ only features terms quadratic, linear  and constant in $p_\varphi$ so that the  integration over $p_\varphi$ can be easily done with the stationary-point method. Including the effect of $\vec\Omega$, the argument of the exponential in~(\ref{GenFun}) is stationary at
  \be p_\varphi = \dot\varphi-\vec\Omega\cdot  \vec x \times \vec\nabla\varphi.  \ee
  Using this expression in integrating over $p_\varphi$ in~(\ref{GenFun}), one finds the following Lagrangian path integral:
   \be\boxed{\mathcal{Z}(j) = \frac1{``j\to 0"} \int \delta\varphi \exp\left(iS_C(\varphi)+i\int_C d^4x j(x)\varphi(x^\omega)\right),}  \label{GenFunLag} \ee
   where the denominator ``$j \to0$" is the numerator for $j \to0$,
   \be \boxed{S_C(\varphi) = \int_C d^4x \left(-\frac12 \varphi (\Box_\omega+M^2)\varphi - \mathcal{V}(\varphi) \right)} \label{SC}\ee
 is the action computed with the contour $C$ and the operator $\Box_\omega$   is the d'Alembertian operator after the substitution $\partial_t \to \partial_t -\vec\Omega\cdot \vec x\times\vec\nabla$.  
 
 It is useful to note that $\vec\Omega$ only appears in the  quadratic action. 
 This implies that, in perturbation theory, only the propagators are modified by $\vec\Omega$, the vertices are unmodified. For the computation of the vertices one can use well-known results from the literature (see e.g.~\cite{Bellac:2011kqa}  and~\cite{Landsman:1986uw});
 on the other hand, the propagators have been computed, including the effect of $\vec\Omega$ and  $\mu_a$,  in Sec.~\ref{Thermal propagator}. This provides a practical recipe to perform perturbation theory in the presence of both $\vec\Omega$ and  $\mu_a$.
 
 Eq.~(\ref{SC}) shows that, as geometrically clear, $\vec\Omega$ breaks rotational and space-translational symmetry down to cylindrical symmetry (with axis in the direction of $\vec\Omega$), while time-translational symmetry is preserved. In cylindrical coordinates~(\ref{CyCoo})
  \be \vec\Omega\cdot \vec x\times\vec\nabla = \Omega \partial_\phi \ee
 and the operator in the quadratic action reads
 \be \Box_\omega+M^2 
 =(\partial_t-\Omega\partial_\phi)^2 - \partial_r^2 -\frac1r \partial_r-\frac1{r^2} 
 \partial_\phi^2 - \partial_z^2 +M^2. \label{Opomega} \ee
 So  the term due to $\vec \Omega$ does not contribute positively to the Euclidean action, which one has to consider in the imaginary-time formalism.

 This situation is reminiscent of the well known sign problem due to the chemical potentials.
 To understand more clearly this point one can consider the following change of variables in the path integral:
 \be d(t,\vec x)\equiv e^{-i\mu_a\theta^a t}\varphi(t,\vec x), \ee
 which can be viewed as the action of a time-dependent element of $\mathcal{G}$ and satisfies, because of~(\ref{PerCon}), $$d(t_0-i\beta,\vec x) = d(t_0,\vec x).$$ As a result, the generating functional in~(\ref{GenFunLag}) is left invariant except the fact that one has to impose this simpler periodicity condition and perform the   following substitutions:
 \be \partial_t \to \partial_t+i\mu_a\theta^a, \qquad j(t,\vec x) \to e^{i\mu_a\theta^a t} j(t,\vec x). \ee
So   $S_C$ in~(\ref{SC})
reads  
\be S_C=\int_C d^4x \left\{ -\frac12d\left[(\partial_t -\Omega\partial_\phi+i\mu_a\theta^a)^2-\vec\nabla^2+M^2\right] d -\mathcal{V}(d) \right\}. \nonumber
\ee
As a result, both $\mu_a$ and $\Omega$ give a contribution to the Euclidean action that is not positive definite (recall that the $\theta^a$ are purely imaginary and antisymmetric and so Hermitian). 

To entirely avoid the issue that this may create in doing a non-perturbative calculation on the lattice,  one can substitute $\mu_a \to i \mu^E_a$ and  $\Omega\to i\Omega_E$, treat $\mu^E_a$ and $\Omega_E$ as real parameters to evaluate the path integral in Euclidean spacetime and finally perform the analytic continuation 
$\mu^E_a\to -i\mu_a$, $\Omega_E\to -i \Omega$ to obtain results of physical relevance. 

It is interesting to note, though, that these substitutions may not be necessary in some cases. Consider, for example, $\mu_a=0$ and rewrite the operator in~(\ref{Opomega}) as
\be \Box_\omega+M^2 =\partial_t^2-2\Omega\partial_\phi\partial_t +\left(- \partial_r^2 -\frac1r \partial_r-\frac1{r^2} 
 \partial_\phi^2 +\Omega^2\partial_\phi^2\right) - \partial_z^2 +M^2. \ee 
 The first term is the familiar $-\partial_\tau^2$ at imaginary time $\tau$ and contributes positively to the Euclidean action. The same is true for $- \partial_z^2$ and $M^2$ if its eigenvalues are non negative.  The second term becomes $-2i\Omega\partial_\phi\partial_\tau$ and gives a harmless imaginary contribution to the  Euclidean action. Finally, note that the operator in the brackets has eigenvalues $\alpha^2-m^2\Omega^2$. While the contribution of $\Omega$ is negative as already stated, in Sec.~\ref{Computing ensemble averages} we have seen, putting the system in a finite cylinder of radius $R$ and height $L$, that $1/R>\Omega$ and also $\alpha\geq |m|/R$~\cite{Vilenkin:1980zv} so that $\alpha^2-m^2\Omega^2>0$, namely a positive contribution to the Euclidean action. In the same section we have also seen that in the large $R$ and $L$ limit, $\alpha$ and $y\equiv m/R$ become continuous variables and $\alpha^2-m^2\Omega^2 = \alpha^2 -v^2y^2>0$, where the positivity is now due to $\alpha\geq|y|$ and also $0\leq v<1$. So one can see that the negative contribution of $\Omega$ is  compensated by other positive contributions. 

The conclusion is, therefore, that the real part of the Euclidean action is bounded from below, at least for $\mu_a=0$. This opens up the possibility to perform non-perturbative calculations on the lattice in the imaginary-time formalism, even in the presence of a non-zero average angular momentum.

\section{An application: weakly-coupled particle production}\label{Applications}

The results in Secs.~\ref{Interacting theories and general path-integral formula} and~\ref{MomInt} give ready-to-use formul\ae~to study arbitrary and generically interacting scalar field theories.

As an application of the formalism developed so far, let us compute  the production of spin-0 particles that are weakly-coupled to a thermal bath, i.e.~a system of other particles in thermal equilibrium, for the most general equilibrium density matrix introduced in Sec.~\ref{density matrix}. The produced particles are not necessarily in thermal equilibrium as their couplings are small.   Here the real-time formalism, in the form discussed in Sec.~\ref{Interacting theories and general path-integral formula}, is used: this is  the most convenient formalism to compute particle interaction rates as one does not have to perform an analytic continuation on time.

The spin-0 particles are described by a real scalar field $h$ of mass $\mu$. In the interaction picture $h$ can be decomposed like in~(\ref{PhiDec}), using its creation and annihilation operators. The interaction between $h$ and the thermal bath is assumed to be given by a term $\lambda h O$ in the Lagrangian, where $O$ is a local and real scalar operator composed by the fields in thermal equilibrium and $\lambda$ is a small coupling constant. Therefore, $h$ may not be in thermal equilibrium with the fields in $O$.

 The $S$-matrix element for such particle  production is (at leading order in $\lambda$)
\be S_{if}(q)  \simeq i\lambda \int d^4x \, \langle f, q|h(x) O(x) |i\rangle = i\lambda \sqrt{\Delta\omega \Delta p} \int d^4x \,\varphi_q^*(\vec x) e^{i\omega t} \langle f|O(x) |i\rangle,\ee
where $|i\rangle$ and $|f,q\rangle$ are the initial and final states, respectively, and the orthonormal basis of eigenstates of the operator $H-\vec\Omega  \cdot \vec J - \mu_a Q^a$, appearing in the density matrix (see Eq.~(\ref{rhoRest})), is chosen. Moreover, $q\equiv \{\omega, p, m\}$ represents the energy $\omega$, the linear momentum $p$ and angular momentum $m$ along the rotation axis of the produced particle. For the time being, the system is 
enclosed in a finite volume for technical convenience; the infinite volume limit will be taken later. The production probability averaged over the initial state and summed over $f$ is
\be  \frac{1}{Z}\sum_{if} e^{-\beta \mathcal{E}_i} |S_{if}(q)|^2 \simeq \frac{\lambda^2\Delta\omega \Delta p}{Z}\sum_{if} e^{-\beta \mathcal{E}_i} \int d^4x_1d^4x_2 \, \varphi^*_q(\vec x_1) \varphi_q(\vec x_2)e^{i\omega(t_1-t_2)} \langle i|O(x_2)|f\rangle\langle f|O(x_1)|i\rangle, \nonumber \ee
where the $\mathcal{E}_i$ are the eigenvalues of $H-\vec\Omega  \cdot \vec J - \mu_a Q^a$. 
Since $\lambda$ is small, i.e.~$h$ is weakly-coupled, one can neglect $\lambda$ in everything that multiplies $\lambda^2$ in the expression above and write
\be\sum_f\langle i|O(y)|f\rangle\langle f|O(x)|i\rangle= \langle i|O(y)O(x)|i\rangle. \ee
So  at leading order in $\lambda$, using also~(\ref{WFcyl}), 
\be \boxed{\frac{1}{Z}\sum_{if} e^{-\beta \mathcal{E}_i} |S_{if}(q)|^2  \simeq \frac{\Delta\omega\Delta p}{2(2\pi)^2} \int d^4x_1d^4x_2 \, J_m(\alpha r_1)J_m(\alpha r_2) e^{iq(\xi_1-\xi_2)}  \mathcal{G}^<(\xi_1-\xi_2, r_1, r_2),} \label{ProbProd}\ee 
where $q(\xi_1-\xi_2)\equiv \omega(t_1-t_2)- p(z_1-z_2) - m (\phi_1-\phi_2)$ and 
\be \mathcal{G}^<(\xi_1-\xi_2, r_1, r_2)\equiv\lambda^2\langle O(x_2)O(x_1)\rangle = \frac{\lambda^2}{Z}\sum_i e^{-\beta \mathcal{E}_i} \langle i|O(x_2)O(x_1)|i\rangle.\ee
In the expression above we have highlighted that $\langle O(x_2)O(x_1)\rangle$ is a function, $\mathcal{G}^<$, of $r_1$, $r_2$ and the difference $\xi_1-\xi_2\equiv \{t_1-t_2, z_1- z_2, \phi_1-\phi_2\}$, not of the sum $\xi_1+\xi_2$, because time translations as well as space translations along the $z$ axis and rotations around the $z$ axis are preserved. $\mathcal{G}^<$ represents the  ``non time-ordered" 2-point function of the operator $O$.

It is important to note that the result in~(\ref{ProbProd}), although only at leading order in $\lambda$, is  valid to all orders (and even non-perturbatively) in the couplings of the thermalized sector other than $\lambda$.

 If perturbation theory holds, one can compute $\mathcal{G}^<$ with the Kobes-Semenoff rules~\cite{KS,KS2}, an extension to TFT of the well-known cutting rules in non-statistical quantum field theory. In~\cite{KS,KS2} Kobes and Semenoff assumed $\vec\Omega = 0$ and $\mu_a = 0$, but, as shown in Sec.~\ref{MomInt}, only the propagators are modified by $\vec\Omega$ and $\mu_a$, the vertices are unmodified. The propagators have been computed, including the effect of $\vec\Omega$ and $\mu_a$,  in Sec.~\ref{Thermal propagator}.
 
Eq.~(\ref{ProbProd}) is a general formula for the production of a spin-0 particle that is weakly coupled to an arbitrary equilibrium plasma. As an example of how~(\ref{ProbProd}) can be used in practice, let us consider the production of a Higgs boson through a  portal coupling $\lambda h S^2$, where $S$ is a real scalar of mass $\mu_s$ belonging to a dark sector ($\lambda$ is supposed to be very small) and $h$ here represents the Higgs field in the unitary gauge. 
 It is useful to have a mechanism to produce Standard Model (SM) particles, such as Higgs bosons, from dark sectors. 
 Indeed, these sectors can feature interesting phenomena, like first-order phase transitions and/or may be responsible for an inflationary expansion of the universe, which, to be viable, would eventually require SM particle production in a reheating epoch. Regarding the topic of the present paper, it is relevant to note that these phenomena could be accompanied by production of compact objects such as primordial black holes, which can feature rotating ($\Omega\neq0$) accretion disks and coronas~\cite{Abramowicz:2011xu,DeLuca:2019buf,DeLuca:2020bjf,Harada:2020pzb,Jaraba:2021ces,Banerjee:2023qya,Harada:2024jxl,Banerjee:2024nkv,Banerjee:2024cwv}.
  
  If an extra sector typically  features energy and mass scales much above the electroweak scale, its couplings to the SM particles are not constrained to be very small by experimental bounds and the SM particle production can easily be sizable. In the complementary situation where  typical energy and mass scales are below the electroweak scale an extra sector must be dark. In this situation the relevant operator $h S^2$ dominates over the marginal operator $h^2 S^2$.
 
\begin{figure}[t]
\begin{center}
  \includegraphics[scale=0.6]{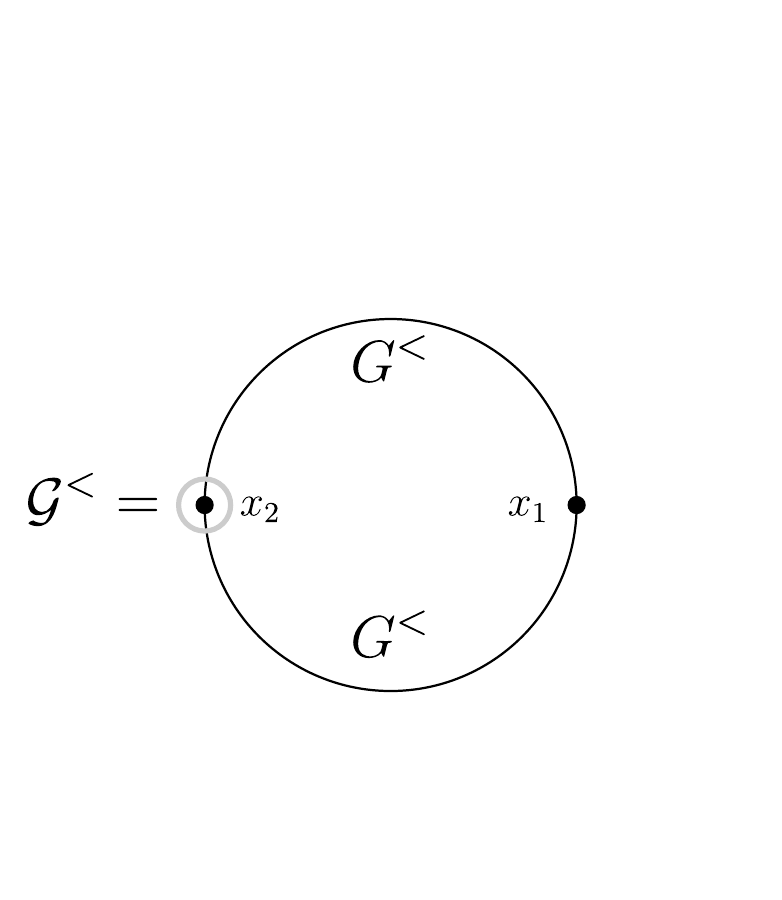} \quad  \includegraphics[scale=0.45]{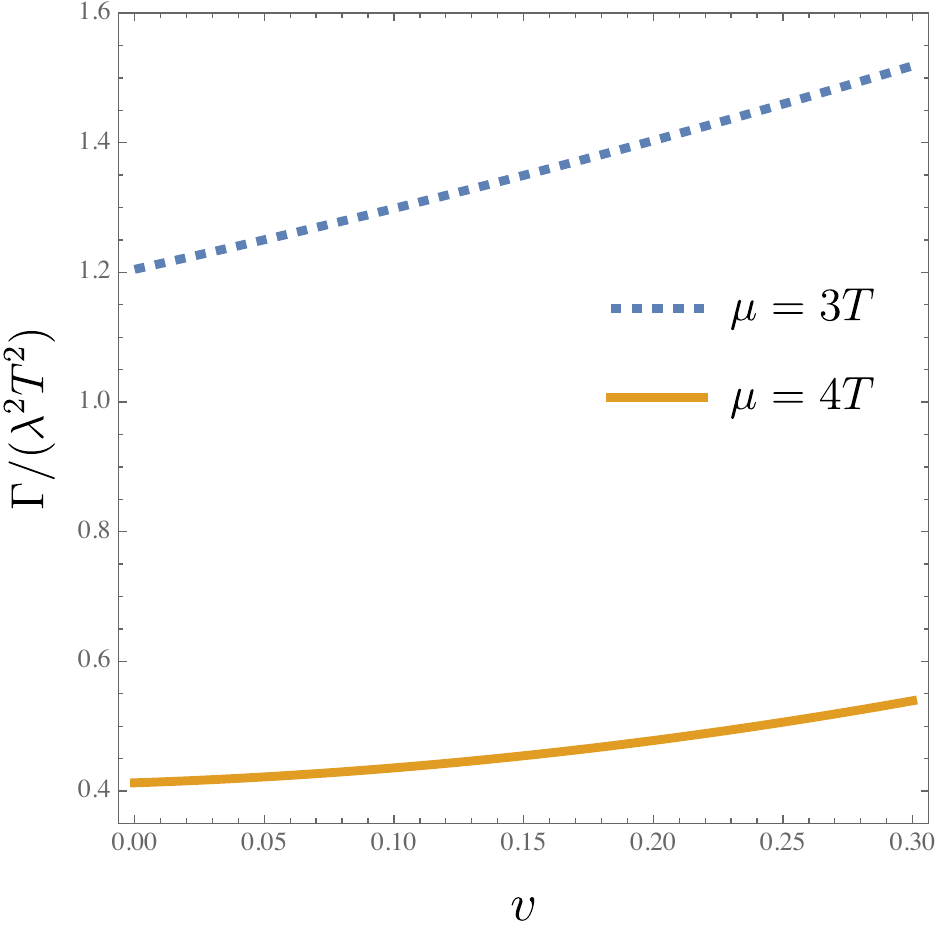}  
    \caption{\em The left plot shows a diagram representing the  ``non time-ordered" 2-point function of the composite operator $O=S^2$ in terms of the  ``non time-ordered" 2-point functions of $S$. The Kobes-Semenoff circling notation~\cite{KS,KS2} is used. The right plot shows the corresponding production rate per unit of volume, $\Gamma$, in Eq.~(\ref{Gamma}) in the large space volume limit for the cubic interaction $\lambda h S^2$,  $\mu_s\ll T$ and some values of the mass $\mu$ of the produced particle.}\label{diagram}
  \end{center}
\end{figure}

 For the above-mentioned portal coupling, applying the  Kobes-Semenoff rules with the $\vec\Omega$-dependent propagators, as described by Fig.~\ref{diagram},  one finds
 \be \langle O(x_2)O(x_1)\rangle = 2 G^<(x_1,x_2)^2, \label{OOG}\ee
where the free function $G^<$ has been computed in\footnote{The factor of $2$ in~(\ref{OOG}) comes from the two possible contractions between different spacetime points, while the  contractions between equal spacetime points do not contribute to~(\ref{ProbProd}) because the energy of the produced particle must not be zero. Indeed, the latter type of  contractions  leads to  a time-independent term in  $\langle O(x_2)O(x_1)\rangle$ and then the integral over time in~(\ref{ProbProd}) gives $\delta(\omega)$, which vanishes for $\omega >0$.}
 Sec.~\ref{Thermal propagator}.
 
For $O=S^2$ (noting that there is no chemical potential for a single real scalar),  one finds
\bea 2 G^<(x_1,x_2)^2 &=& \SumInt_{q_1,q_2} \frac{J_{m_1}(\alpha_1 r_1)J_{m_2}(\alpha_2 r_1)J_{m_1}(\alpha_1 r_2)J_{m_2}(\alpha_2 r_2)}{2(2\pi)^4}\left[e^{i(q_1+q_2)(\xi_1-\xi_2)} (1+f_B^{q_1})(1+f_B^{q_2})\right. \nonumber \\
&&\left. +2e^{i(q_1-q_2)(\xi_1-\xi_2)} (1+f_B^{q_1})f_B^{q_2}+ e^{-i(q_1+q_2)(\xi_1-\xi_2)} f_B^{q_1}f_B^{q_2} \right], \label{stepGamma} \eea 
 with $\alpha_i \equiv \sqrt{\omega_i^2- p_i^2 - \mu_s^2}$ and $f_B^{q_i} \equiv f_B(\omega_i - m_i \Omega)$.  Inserting this expression in~(\ref{ProbProd}) leads to the loop integral\footnote{The first term in~(\ref{stepGamma}) proportional to $(1+f_B^{q_1})(1+f_B^{q_2})$ does not contribute because one should evaluate it at $q_1+q_2 =-q$, which has no solution as $\omega>0$, $\omega_1>0$ and $\omega_2>0$.}
 \be \frac{1}{Z}\sum_{if} e^{-\beta \mathcal{E}_i} |S_{if}(q)|^2 \simeq \frac{\lambda^2\Delta\omega\Delta p \Delta t L}{4(2\pi)^2}\SumInt_{q_1}\left[I^2_{q,q_1,q-q_1} f_B^{q_1}f_B^{q-q_1} +2 I^2_{q,q_1,q+q_1} (1+f_B^{q_1})f_B^{q+q_1}\right], \label{ProbLoop}\ee
 where $\Delta t$ is the full time interval and the $I_{q,q_1,q_2}$ represent the radial integrals
 \be I_{q,q_1,q_2} \equiv \int_0^\infty dr r J_m(\alpha r)J_{m_1}(\alpha_1 r)J_{m_2}(\alpha_2 r).  \ee 
 
 The second contribution in~(\ref{ProbLoop}) must vanish as it represents the $h$ emission from an on-shell $S$, which, if not vanishing, would violate energy-momentum conservation. 
 To see it explicitly note that $I_{q,q_1,q_2}$ vanishes if $\alpha$, $\alpha_1$ and $\alpha_2$ do not form the sides of a triangle~\cite{JacksonMaximon}. Requiring that they do form the sides of a triangle one can define three bi-dimensional vectors $\vec\alpha$, $\vec\alpha_1$ and $\vec\alpha_2$ such that $\vec\alpha_2=\vec\alpha+\vec\alpha_1$. Then one groups together in three four-vectors $\{\omega, \vec\alpha, p\}$, $\{\omega_1, \vec\alpha_1, p_1\}$ and $\{\omega_2, \vec\alpha_2, p_2\}$. If the $S$ particle is massless it cannot decay, if it is massive, going to the rest frame of the particle 2 one would violate energy conservation. 
 
 The first contribution in~(\ref{ProbLoop}) represents the $h$ production through coalescence of two $S$  particles and does not vanish at finite temperature. It is important to note that the loop integral in~(\ref{ProbLoop}) is performed on an integration domain such that $q_2$ (which equals $q-q_1$ in the first non-vanishing contribution) corresponds to a real particle.

 To compute~(\ref{ProbLoop}) one only needs the integrals $I_{q,q_1,q_2}$ for values of the integer $m$, $m_1$ and $m_2$ such that $m+m_1+m_2=0$ because the cylindrical Bessel functions satisfy $J_{-m}(x) = (-1)^mJ_m(x)$ for integer values of $m$~\cite{AbramowitzStegun} and $I_{q,q_1,q_2}$ only appears quadratically in~(\ref{ProbLoop}). For $m+m_1+m_2=0$ the $I_{q,q_1,q_2}$ are known analytically~\cite{JacksonMaximon}: if $\alpha$, $\alpha_1$ and $\alpha_2$ do not form the sides of a triangle then $I_{q,q_1,q_2} =0$, otherwise
 \be I_{q,q_1,q_2} = \frac{\cos(m\gamma_1-m_1\gamma)}{2\pi A}, \ee
 where the angles $\gamma$ and $\gamma_1$ are given by
 \be \cos\gamma=\frac{\alpha^2-\alpha_1^2-\alpha_2^2}{2\alpha_1\alpha_2}, \qquad \cos\gamma_1 = \frac{\alpha_1^2-\alpha^2-\alpha_2^2}{2\alpha\alpha_2}, \ee 
 and $A$ is the area of the triangle of sides $\alpha$, $\alpha_1$ and $\alpha_2$, which is given by Heron's formula:
 \be A=\frac14\sqrt{(\alpha+\alpha_1+\alpha_2)(\alpha_1+\alpha_2-\alpha)(\alpha+\alpha_2-\alpha_1)(\alpha+\alpha_1-\alpha_2)} .\ee 
 These formul\ae~allow us to compute the production probability~(\ref{ProbLoop}).

 In particular, one is interested in computing the full $h$-production rate $\Gamma$ per unit of
 volume $\pi R^2 L$, which is obtained by summing~(\ref{ProbLoop}) over all possible values of $q$ (and dividing by $\pi R^2 L$)
 \be \Gamma = \frac{\lambda^2}{16\pi^3 R^2}\SumInt_{q, q_1}I^2_{q,q_1,q-q_1} f_B^{q_1}f_B^{q-q_1}. \label{Gamma}\ee 
 
 Note that $\Gamma$ has a remarkable property: it grows indefinitely when the velocity parameter $v \equiv \Omega R$ approaches 1. This can be seen by noting that the convergence of the sum over $m$ in~(\ref{Gamma}) requires
\be \Omega <  \lim_{m\to +\infty}\frac1{m}\min_{n\,p} ( \omega_{m,n}(p)-\omega_{m_1,n_1}(p_1) +m_1\Omega) = \lim_{m\to +\infty} \frac{j_{m,1}}{m R} = \frac1{R}.  \label{ConvCond2} \ee 
Such property can be used to enhance the  rate of particle production through a rotating plasma with a sizable value of $v$. 
 
Eventually one is interested in taking the $L\to\infty$ and $R\to\infty$ limit where the full space is recovered, to remove any dependence on the shape of the finite-volume region. No infrared divergences are present in this limit. To show this one can substitute  $1/R^2 \to dy dy_1$ because $y \equiv m/R$ and $y_1 \equiv m_1/R$ become continuous variables in this limit and $m$ and $m_1$ are integers. As we have seen in 
Sec.~\ref{Computing ensemble averages} around Eq.~(\ref{Deltaalpha}), $y$ and $y_1$ vary respectively in the intervals $[-\alpha, \alpha]$ and $[-\alpha_1, \alpha_1]$.   With these manipulations $\Gamma$ can be written as an integral over six variables of an integrand that never diverges as fast as the inverse of a monomial of degree six in the integration variables (because the produced particle, such as the Higgs boson, must be massive to be produced through coalescence).  As a result $\Gamma$ is free from infrared divergences.

As we have shown, $\Gamma$ assumes very large values as $v$ approaches 1. For moderate values of $v$ the dependence, computed numerically, is shown in the right plot of  Fig.~\ref{diagram} for some benchmark values of the $h$ mass, $\mu$. Already for small $v$ one can note a sizable growth of $\Gamma$ increasing $v$.

\section{Summary and conclusions}\label{Conclusions}

Let us conclude by providing a summary of the main original results obtained.
\begin{itemize}
\item Sec.~\ref{density matrix} showed (in Sec.~\ref{Rest frame}) how to express the most general equilibrium density matrix in a simple frame where the plasma is at rest, Eqs.~(\ref{rhoRest0}) and~(\ref{rhoRest}). This can be done by performing a specific space translation followed by a specific Lorentz transformation. After these transformations are performed, the density matrix is a function of commuting operators only, which significantly simplifies the subsequent analysis. However, by inverting the above-mentioned transformations one can always recover the expressions in a generic inertial frame. In Sec.~\ref{Ensemble averages} the general expressions for  the averages of $H$, $\vec J$ and $Q^a$ are presented in terms of the partition function. These expressions can be combined with the results of the subsequent sections to obtain those averages in an arbitrary scalar theory (with or without interactions). It is worth noting, though, that the results of Sec.~\ref{density matrix} apply to fields of any spin.
\item In Sec.~\ref{Free fields} the analysis of TFT for a generic equilibrium density matrix started with the simplest case of non-interacting particles. However,~\ref{Free fields} included the most general spin-0 particle content, featuring generic masses, chemical potentials and thermal vorticity (corresponding to the average angular momentum). The cylindrical coordinates with axis along the thermal vorticity were adopted to exploit the residual space symmetries. Moreover, the system was initially enclosed in a cylinder of height $L$ and radius $R$ for computational convenience.  

In Sec.~\ref{Computing ensemble averages} the averages of the product of two annihilation and creation operators were derived in a closed form, which allowed us to compute the averages of $H$, $\vec J$ and $Q^a$. An important finding is that the convergence of the averages requires $\Omega<1/R$. The large-volume limit can be taken by keeping $v\equiv R\Omega\in[0,1)$ fixed, but that convergence requirement implies that the averages $\langle \rho_E\rangle$, $\langle\mathcal{J}_z\rangle$ and $\langle \rho_a\rangle$ in~(\ref{rhoEp}),~(\ref{CallJz}) and~(\ref{avQcon}) increase indefinitely as 
$v$ approaches $1$. Thus, by varying $v\in[0,1)$ at fixed temperature and chemical potentials one can obtain all values of the average angular momentum, then the average energy and charges are predicted. Moreover, in~(\ref{BoundChem}) bounds on the chemical potentials in terms of $v$ and the mass $\mu$ of a particle in a generic  irreducible representation were found. These bounds are novel generalizations to arbitrary $v\in[0,1)$ (and arbitrary $\mu_a$) of a well-known bound.  One might expect that the results of Sec~\ref{Computing ensemble averages} can be trivially extended to bosons of any spin.

In Sec.~\ref{Thermal propagator} the thermal propagator and the ``non-time-ordered" 2-point functions of a field in a generic irreducible representation with arbitrary chemical potentials and thermal vorticity was obtained. This was done by exploiting the averages of the product of two annihilation and creation operators previously calculated. Just like in non-statistical quantum field theory, the thermal propagator is an important ingredient to perform perturbation theory in an interacting perturbative theory. 
\item The description of a generically interacting theory was then provided in Sec.~\ref{Interacting theories and general path-integral formula} by deriving path-integral expressions for the partition function, Eq.~(\ref{GenPart}), and the thermal Green's functions, Eqs.~(\ref{PIscalars}). A formalism which includes the real- and imaginary-time formalism was adopted, by generalizing existing methods to include the average angular momentum. 
In the future this formalism may perhaps be generalized to non-equilibrium density matrices using Schwinger-Keldysh methods, see e.g.~\cite{Barvinsky:2023jkl}.
\item In order to obtain formul\ae~that can be directly used in cases of interest, in Sec.~\ref{MomInt} the momentum integration in the path integral was performed in the case of ``ordinary scalars", i.e. scalar theories with Hamiltonian of the form~(\ref{Hordinary}). This led to the familiar Lagrangian path integral, where, however, the effect of $\Omega$ is also included. In the same section it was shown that the $\Omega$-dependent terms, like the $\mu_a$-dependent ones, do not contribute positively to the Euclidean action. In order to study non perturbatively a TFTs with  non-vanishing $\Omega$ and $\mu_a$ one can define the Euclidean theory starting with the formal substitution $\Omega\to i \Omega_E$, $\mu_a\to i \mu_a^E$, treat  $\Omega_E$ and $\mu_a^E$ as real parameters in the non-perturbative calculations and finally perform the analytic continuation $\Omega_E \to -i \Omega$ and $\mu_a^E \to - i \mu_a$ (together with $\tau\to it$). However, we have seen that these substitutions  may not be always necessary. For example, in the case $\mu_a=0$, while $\Omega$ does not contribute positively to the Euclidean action, its negative contributions can be compensated by other contributions that are instead positive. 
\item Sec.~\ref{Applications} provided some applications of the previously obtained results. One straightforward application is the definition of an effective action along the lines of~\cite{Salvio:2024upo}, which could be used in the future, for example, to study phase transitions for a general equilibrium density matrix. In this paper we focused on another application:  the production of a spin-0 particle weakly-coupled to a rotating thermal plasma. In a first part of that section a general expression for the production probability was obtained, Eq.~(\ref{ProbProd}). That result is valid for arbitrary mass of the produced particle and arbitrary temperature, chemical potentials and average angular momentum of the plasma. Furthermore, it is valid even if the plasma is composed by particles of any spin. Eq.~(\ref{ProbProd}) could then be used to compute the production rate of a weakly coupled spin-0 particle in an arbitrary model, for example, Higgs bosons and pions in the SM or axions~\cite{Weinberg:1977ma}
 in any Peccei-Quinn model~\cite{Peccei:1977hh}; this would extend the calculations in~\cite{Salvio:2013iaa,DEramo:2021psx,Arias-Aragon:2020shv,Iwamoto:1984ir,Springmann:2024mjp} performed in a non-rotating plasma.
As a concrete example, Eq.~(\ref{ProbProd}) was then applied to compute the production rate of a Higgs boson coupled to  a dark sector through a portal scalar coupling. Interestingly, it was found that the production rate can be significantly increased by $v$. The formula obtained, Eq.~(\ref{Gamma}), could be used, for example, to improve the production of SM particles after an inflationary expansion of the universe due to the dynamics of a dark sector, provided that a mechanism to generate a rotating plasma is at work. A mechanism of this sort could be due to the presence of rotating primordial black holes, generating rotating accretion disks and coronas.  
 \end{itemize}
 
 \vspace{0.7cm}
 
 \subsection*{Acknowledgments}
I thank Francesco Tombesi for valuable discussions on rotating plasmas around black holes.

\vspace{1cm}

 \footnotesize
\begin{multicols}{2}

\end{multicols}

  \end{document}